\documentclass[aps,adv,twocolumn,superscriptaddress]{revtex4-2}
\usepackage{epsfig}
\usepackage{graphicx}
\usepackage{dcolumn}
\usepackage{bm}
\usepackage{latexsym}
\usepackage{color}
\usepackage{amsmath}
\usepackage{amssymb}
\usepackage{cleveref}
\usepackage{mathtools}
\usepackage{tikz}
\usepackage{pgfplots}
\usepackage{xcolor,colortbl}

\definecolor{Gray}{gray}{0.90}
\definecolor{DarkSand}{gray}{0.70}

\newcommand{\andd}{\quad\mbox{and}\quad}

\allowdisplaybreaks[1]

\pgfplotsset{compat=1.18}
\begin{document}

	 \title{Pearcey integrals, Stokes lines and exact baryonic layers\\ in the low energy limit of QCD}
	
	\author{Sergio L. Cacciatori}
	
	\affiliation{Universit\`a  dell'Insubria, Dipartimento di Scienza ed Alta Tecnologia, via Valleggio 11, 22100, Como, Italy}
	\affiliation{INFN, via Celoria 16, 20133, Milano, Italy}

	\author{Fabrizio Canfora}

    \affiliation{Universidad San Sebastián, Facultad de Ingeniería, Arquitectura y Diseño, sede Valdivia, General Lagos 1163, Valdivia 8420524, Chile}
	\affiliation{Centro de Estudios Cient\'{\i}ficos (CECS), Casilla 1469, Valdivia, Chile}
	
	\author{Federica Muscolino}
	 \affiliation{Universidad San Sebastián, Facultad de Ingeniería, Arquitectura y Diseño, sede Valdivia, General Lagos 1163, Valdivia 8420524, Chile}
	\affiliation{Centro de Estudios Cient\'{\i}ficos (CECS), Casilla 1469, Valdivia, Chile}
	
%
	
	\begin{abstract}
		The first analytic solutions representing baryonic layers living at finite baryon density within a constant magnetic field in the gauged Skyrme model are constructed. A remarkable feature of these configurations is that, if the Skyrme term is neglected, then these baryonic layers in the constant magnetic background cannot be found analytically and their energies grow very fast with the magnetic field. On the other hand, if the Skyrme term is taken into account, the field equations can be solved analytically and the corresponding solutions have a smooth limit for large magnetic fields. Thus, the Skyrme term discloses the universal character of these configurations living at finite Baryon density in a constant magnetic field. The classical gran-canonical partition function of these configurations can be expressed explicitly in terms of the Pearcey integral. This fact allows us to determine analytically the Stokes lines of the partition function and the corresponding dependence on the baryonic chemical potential as well as on the external magnetic field. In this way, we can determine various critical curves in the ($\mu_B-B_{ext}$) plane which separates different physical behaviors. These families of inhomogeneous baryonic condensates can be also dressed with chiral conformal excitations of the solutions representing modulations of the layers themselves. Some physical consequences are analyzed.
	\end{abstract}

	\maketitle
	\tableofcontents
	

\section{Introduction}


Non-perturbative effects of Quantum Chromodynamics (QCD) play a fundamental role in the low-energy processes of strong interactions since the color confinement prevents from describing the low energy limit of systems of quarks through perturbative approaches \cite{Greensite:LectureNotesConfinement,Ripka:LecturesDualSuperconductors,Shifman1}. The unknown behavior of strong interactions in this regime makes the understanding of certain phenomena a hard task; it is the case, for instance, of the internal structure of Neutron Stars. Indeed, the intermediate layers of these objects are known to be composed by particular states of baryonic matter, called \textit{Nuclear Pasta} \cite{HorowitzSchneider:2013dwa,Horowitz:2014xca,Newton:2021vyd,Newton:Nature,Pons:2013nea,Lopez:2020zhk,Dorso:2020zhk,Klebanov:1985qi}. Here, baryons appear to be deformed and organized in ordered arrays with different shapes, which reminds of pasta (for instance, they can form tubes, which reminds of \textit{spaghetti states}, or layers, which reminds of \textit{lasagna states}, or balls, which remind of \textit{gnocchi states}, and so on). Their formation is due to the competition of non-perturbative QCD effects and Coulomb interactions. An analytical description of these interactions is, by now, out of reach; for this reason, they are generally studied employing numerical simulations \cite{HorowitzSchneider:2013dwa,Horowitz:2014xca,Horowitz:Saturation,Horowitz:Elasticity,Newton:Nature,Newton:2021vyd}. With these methods, it is possible to characterize the general structures of nuclear pasta and some of its physical properties. The problem of the numerical simulations lies in the fact that the results are limited by the necessity of a huge computational power; for instance, the introduction of external interactions and the study of these nuclear structures in curved space-time is a hard task (see, for instance, \cite{Horowitz:Elasticity}). Alternative methods require a more \textit{fundamental analysis} of the non-perturbative aspects of QCD. An example is provided by the Lattice QCD (LQCD), in which the quark fields are represented by lattice sites and gluons by the links between them \cite{Alford:1998Lattice,Kogut:2002Lattice,Kogut:2002Lattice2,Kogut:2004Lattice,Beane:2007Lattice,Detmold:2008Lattice,Detmold:2008Lattice2,Detmold:2011Lattice,Detmold:2012Lattice,Endrodi:2014Lattice}. When the lattice is infinitely large, the non-perturbative effects arise. As shown in the cited works, LQCD leads to remarkable results, but the non-linear nature of QCD prevents an analytical description; moreover, the numerical results yield the so-called \textit{sign problem}. For this reason, it is not very effective in the study of the phase diagrams of QCD at finite temperatures.

Recent results show that an analytical description of structures of baryons is possible, considering the non-perturbative effects of QCD as properties of some topological spaces; this link is a known fact that has been studied for decades by many authors \cite{Weinberg:ClassicalSoltions,tHooft:UnifiedTheories,Mandelstam:VorticesQuaarkConfinement,Manton:TopologicalSolitons,Coleman:SineGordonThirringModel1974,Witten:CurrentAlgebraGen,Witten:CurrentAlgebraBaryonConfinement}. A remarkable example is represented by the Skyrme model, which has been shown to represent an Effective Field Theory (EFT) for strong interactions, providing a good description of both single baryons and multi-baryonic systems \cite{Skyrme1,Skyrme2,SkyrmeI,SkyrmeIII,SkyrmeIII+,tHooft1NExp,WittenBaryons1NExp,ANW,Canfora1:HedgehogAnsatz,Canfora2:NonlinearSuperposition,Canfora3:ChimPot,Canfora4:AdSWormholes,Canfora5:4dEinstein-nonlinear,Canfora6:U1gauged,Canfora7:Ordered(2018),Canfora8:Lasagne,Canfora9:AnalyticChristals,Canfora10:TraversableNUT-AdSWormholes,Canfora11:SuperCond,Canfora12:Pion,Canfora13:Gauged(May2021),Canfora14:YM,SergioFabrizio:2020zui,NostroI,NostroII}. The main difficult encountered in this representation is represented by the complicated shape of the equations of motion (known as the \textit{Skyrme equations}), which requires the introduction of an ansatz in order to obtain analytical solutions.

Our interest focuses on nuclear pasta structures description; in this scope, we refer to the works \cite{Canfora1:HedgehogAnsatz,Canfora2:NonlinearSuperposition,Canfora3:ChimPot,Canfora4:AdSWormholes,Canfora5:4dEinstein-nonlinear,Canfora6:U1gauged,Canfora7:Ordered(2018),Canfora8:Lasagne,Canfora9:AnalyticChristals,Canfora10:TraversableNUT-AdSWormholes,Canfora11:SuperCond,Canfora12:Pion,Canfora13:Gauged(May2021),Canfora14:YM,SergioFabrizio:2020zui,NostroI,NostroII}, in which two particular ansätze have been introduced. The main goal of these ans\"atze is to reduce the Skyrme equations to one ODE, which allows describing the properties of these baryonic structures through analytical tools. They are based on two parameterizations of the flavor group $G$; namely, the \textit{exponential parameterization} and the \textit{generalized Euler parametrization} (for more details on the generalized Euler parameterization see, for instance, \cite{Tilma:2002ke,Cacciatori:SUN,Cacciatori:EulerGen} and references therein). The employment of the first parameterization is shown to describe a system in which the energy and the baryonic charge concentrate in tubes and, for this reason, are called \textit{baryonic tubes}. Their shape can be reconducted to the nuclear spaghetti states; therefore, the aforementioned ansatz is called \textit{spaghetti ansatz}. The second is suitable for the description of \textit{baryonic layers}, which shape is conductible to the nuclear lasagna states; as for the previous case, the ansatz obtained using the Euler parameterization will be called \textit{lasagna ansatz}. In \cite{Canfora15:CFT}, these ansätze have been extended showing that they can be used to describe the thermodynamical characteristics of nuclear pasta, obtaining interesting differences between the Non-Linear Sigma model (NL$\sigma$M), which is obtained from the Skyrme model omitting the Skyrme term, and the Skyrme model. More specifically, the NL$\sigma$M allows for conformal solutions; whereas, in the Skyrme model, \textit{chiral} conformal solutions are founded. This is a remarkable result since it allows studying the thermodynamics of these baryonic structures from the computation of the partition function.

The techniques to construct analytic inhomogeneous condensates in the above references also work when the Skyrme model is minimally coupled to Maxwell field \cite{NostroI,NostroII}. The exact solutions describing baryonic layers and the tube together with the corresponding Maxwell fields generated by the U(1) Skyrme current have been found by answering the following question: how should one choose the ansatz for the Maxwell potential to keep the solvability properties of the ungauged ansatz in a sector with non-zero Baryonic charge?  This question has a unique answer discussed in \cite{NostroII}. These exact solutions of the gauged Skyrme - Maxwell system are Baryonic layers and tubes together with their electromagnetic field. However, there are important cases (from neutron stars \cite{Kaspi:Magnetars} to heavy Ions collisions \cite{Kharzeev:IonsCollisions,dEnterria:IonsCollisions}) in which the external electromagnetic fields are much stronger than the electromagnetic fields generated by the inhomogeneous Baryonic condensates. When this happens, it is natural to ask:
how intense external electromagnetic fields deform the inhomogeneous Baryonic condensates of the low energy limit of QCD?
It could appear that when the electromagnetic field is an external fixed background field, it should be easier to handle the problem of finding inhomogeneous condensates. Indeed, one may think that the problem of analyzing only the gauged Skyrme field equations in the given background field is easier than the problem (analyzed and solved in \cite{NostroII}) of constructing analytic baryonic layers and tubes together with their electromagnetic fields generated in a self-consistent way. Unfortunately, this is not true as in those references \cite{NostroII} the ansatz for the electromagnetic field is fixed uniquely by requiring a force-free condition \cite{NostroII}. On the other hand, when there is a strong external electromagnetic field, it is necessary to adapt the ansatz for the SU(2)-valued Skyrme field to the external fixed electromagnetic field (which, in general, will not be force-free) and not the other way around. Consequently, a relevant goal is to generalize the approach in \cite{NostroII} in order to describe inhomogeneous Baryonic condensates in strong external electromagnetic fields. The fact that the solution to this technical problem is very far from obvious becomes quite clear if one observes that, until now, no exact analytic Baryonic condensates in an external constant magnetic field has been found.
Therefore, one of the main goals of the present paper is to find explicitly inhomogeneous Baryonic condensates living within strong external electromagnetic fields.

The present work aims to study the behavior of baryonic structures able to describe the nuclear pasta behavior when an external magnetic field is applied. This analysis is motivated by the relevance in astrophysics of the interactions between strong magnetic fields and baryonic structures, such as in the Magnetars \cite{Esposito:Magnetars,Mereghetti:Maagnetars}. We considered the \textit{spaghetti and lasagna ans\"atze} in both the NL$\sigma$M model and the Skyrme model in order to depict the role of the Skyrme term when the soliton is exposed to external solicitations. Indeed, it is shown that the equations in NL$\sigma$M lose coercivity when the external field becomes big \cite{Necas:Coer,Benci1:Coer}. Nevertheless, when the Skyrme term is added, a solution can be always defined, at least in the case of baryonic layers. In particular, in the latter, the solution can be found using analytical tools. A remarkable property of this solution consists in its independence on the value of the external field; for this reason, we are going to call it as \textit{universal solution}.

A remarkable byproduct of our analysis is that we can compute exactly the classical gran-canonical partition function associated with these families of Baryonic layers in a constant magnetic field in terms of the Pearcey integral. The Pearcey integral is well known in optics and has been studied in detail in connection with the Stokes phenomenon and resurgence [cite]. Thus, our exact solutions allow us to compute exactly the Stokes lines of the gran-canonical partition functions in the $\mu_B-B^{\mbox{ext}}$ plane. Taking into account the difficulties of lattice QCD when dealing with the Baryonic chemical potential, the present analytic results which provide explicit expressions for critical lines separating different physical behaviors are especially important.


\subsection{Notation and characteristics of the model}


Let us introduce the main, general characteristics of the considered system. As anticipated in the previous sections, our model is based on the definition of particular ans\"atze which allows us to solve analytically the Skyrme equations for a multi-baryonic system at finite density. The last requirement is satisfied whether the system is confined in a finite space volume (say, a box). Following the notation introduced in \cite{NostroI,NostroII}, the metric is described by
\begin{align}\label{Metric}
	ds^{2}=-dt^{2}+L_{r}^{2}dr^{2}+L_{\theta }^{2}d\theta ^{2}+L_{\phi
	}^{2}d\phi ^{2}.
\end{align}
Here, the variables $r$, $\theta$ and $\phi$ represent dimensionless space variables, with ranges
\begin{equation}
	0\leq r\leq 2\pi \ ,\quad 0\leq \theta \leq \pi \ ,\quad 0\leq \phi \leq
	2\pi \ . \label{PiRanges}
\end{equation}
The spatial dimensions are collected in the quantities $L_i$, which take constant values and define the size of the box.

The interaction with an external $U(1)$ field is described by the Lagrangian
\begin{align}  \label{ActionMaxwell}
	L= \mathrm{Tr} \left[\frac{K}{2}\left(\hat{%
		\mathcal{L}}_{\mu}\hat{\mathcal{L}}^{\mu}+\frac{\lambda}{8}\hat{G}_{\mu\nu}%
	\hat{G}^{\mu\nu}\right)\right],
\end{align}
where
\begin{align}\label{HattedCurrent}
	\hat{\mathcal{L}}_{\mu}=\mathcal{L}_\mu-A_{\mu}^{\mbox{ext}}%
	U^{-1}\left[T,U\right],\ \hat{G}_{\mu\nu}=\left[\hat{%
		\mathcal{L}}_{\mu},\hat{\mathcal{L}}_{\nu}\right].
\end{align}
$U$ represents the Skyrme field, which is given by a map 
\begin{align}
	U:\mathbb{R}^{3+1}\rightarrow G,
\end{align}
where $G$ is a simple, compact Lie group, describing the underlining flavor group of quarks. $\mathcal{L}_\mu$ is the \textit{ungauged} left current, which takes the form 
\begin{align}
	\mathcal{L}_\mu=U^{-1}\partial_{\mu}U=\sum_{i=1}^{dim(G)}\mathcal{L}_{\mu}^iT_i,\ T_i\in\mathfrak{g}.
\end{align}
Here $\mathfrak{g}$ stands for the Lie algebra associated with the flavor group $G$ and $T_i$ are the names of the generators of $\mathfrak{g}$. When $G\equiv SU(2)$, the Lie algebra is denoted by $\mathfrak{su}(2)$ and $T_i$ are represented by the Pauli matrices. Notice that when $A_\mu^{\mbox{ext}}\rightarrow 0$, the Lagrangian matches the ungauged one. The external field is defined along a specific direction in the Lie algebra of the flavor group $G$, given by the element $T\in\mathfrak{g}\equiv Lie(G)$. Since the source of the coupled field is considered to be external to the system, the value of $A_\mu^{\mbox{ext}}$ does not undergo the equations of motions (or, better, it can be interpreted as a solution of some Maxwell equations). In particular, we are interested in the behavior of nuclear pasta in the presence of a constant, strong magnetic field. In this scope, the potential $A_\mu^{\mbox{ext}}$ is taken as a linear function of the space variables with $A_0^{\mbox{ext}}=0$. This way, the electric and magnetic fields are defined respectively by
\begin{align} 
	E^i=F^{0i}=0\andd B^i=\frac{1}{2}\epsilon^{ijk}F_{jk}=const., \label{BConst}
\end{align} 
where $F_{\mu\nu}=\partial_\mu A_\nu^{\mbox{ext}}-\partial_\nu A_\mu^{\mbox{ext}}$ is the field strength. As we are going to analyze in the following sections, the Skyrme equations are further simplified when the magnetic field undergoes some particular conditions. These conditions must be specified case by case.

\subsubsection{The properties of the gauged Skyrme model}

The contribution of the external field to the Skyrme equations takes the general form
{
\begin{align}\label{SkyrmEq}
	\nabla_\mu\left(\hat{\mathcal{L}}^\mu-\frac{\lambda}{4}\left[\hat{G}^{\mu\nu},\hat{\mathcal{L}_\nu}\right]\right)=0,
\end{align}
where $\nabla_\mu$ represents the covariant derivative 
\begin{align}
 \nabla_\mu V^\mu=&\partial_\mu V^\mu +[\mathcal{L}_\mu-A^{\rm ext}_\mu U^{-1}T_3 U,V^\mu]\cr
 =&\partial_\mu V^\mu +[\hat{\mathcal{L}}_\mu-A^{\rm ext}_\mu T_3,V^\mu].
\end{align}
}
We know from \cite{Canfora1:HedgehogAnsatz,Canfora2:NonlinearSuperposition,Canfora3:ChimPot,Canfora4:AdSWormholes,Canfora5:4dEinstein-nonlinear,Canfora6:U1gauged,Canfora7:Ordered(2018),Canfora8:Lasagne,Canfora9:AnalyticChristals,Canfora10:TraversableNUT-AdSWormholes,Canfora11:SuperCond,Canfora12:Pion,NostroI,NostroII} that the solutions to the ungauged case can be found through the \textit{pasta ans\"atze}, which are defined in such a way that the Skyrme equations reduce to one ODE. Now, a question arises spontaneously: Is it possible to find similar (or equal) ans\"atze that preserve the baryonic tubes or layers shapes also when the external field is present? As we will see in the following sections, this question does not have a trivial answer; in particular, when the treated case concerns the baryonic tubes. 

The difficulty of analyzing how the external field affects the shape of nuclear pasta has been already emphasized in the introduction. In this scope, it is necessary to study the behavior of the energy and baryonic densities. 

The energy-momentum tensor is defined in the usual way (see, for instance, \cite{NostroI,NostroII}), substituting the \textit{uncoupled} quantities with the hatted ones. Namely,
\begin{align}
	T_{\mu \nu } &=  -\frac{K}{2}\text{Tr}\biggl(\hat{\mathcal{L}}_{\mu }\hat{\mathcal{L}}%
	_{\nu }-\frac{1}{2}g_{\mu \nu }\hat{\mathcal{L}}_{\alpha }\hat{\mathcal{L}}^{\alpha }\cr
	&\qquad\quad+%
	\frac{\lambda }{4}(g^{\alpha \beta }\hat G_{\mu \alpha }\hat G_{\nu \beta }-\frac{1}{4}%
	g_{\mu \nu }\hat G_{\alpha \beta }\hat G^{\alpha \beta })\biggl) ,  \label{TmunuSUN}
\end{align}
from which the energy density is derived ($\rho_E=T_{00}$). 

The baryon charge is computed using
\begin{gather}  \label{rhoB}
	B=\int_{\mathcal{V}} \rho_B,\quad\mbox{with}\quad\rho_B=\frac{1}{24\pi^{2}} \mathrm{Tr} (\mathcal{L }\wedge 
	\mathcal{L }\wedge \mathcal{L}),
\end{gather}
where $\mathcal{V}$ is the spatial region spanned by the coordinates at any
fixed time $t$. Here, a comment is in order. In \cite{NostroII} a general definition of the gauge invariant baryonic charge is given, and it involves a non-trivial contribution of the gauge field, namely,
\begin{align}
	\hat \rho_B&=\rho_B+3 \varepsilon^{ijk} \partial_i [A_j^a \mathrm{Tr}(%
	\mathcal{L}_k (T_a+U^{-1} T_a U))\cr
	&\qquad\qquad\qquad-A^a_j A^b_k \mathrm{Tr} (T_a U^{-1} T_b
	U)].
\end{align}
However, in the present paper $A_\mu^{\mbox{ext}}$ is a fixed external background, so it cannot be included in the definition of the topological baryonic charge of the Skyrmion. For this reason, for the gauged baryonic density we take $\hat{\rho}_B=\rho_B$. However, notice that the external field contribution still enters through the solution to the equations of motion.

\subsubsection{The Non-Linear Sigma Model vs. the Skyrme model}
We underline the importance of the Skyrme term, which defines the difference between the NL$\sigma$M (which is represented by the Skyrme model with $\lambda=0$) and the Skyrme model. The Skyrme term was originally introduced to stabilize the topological structure of a single soliton defined in $\mathbb{R}^3$ \cite{Skyrme1,Skyrme2,SkyrmeI,SkyrmeIII,SkyrmeIII+}. As studied in \cite{NostroI}, this phenomenon is also observed when the structures of baryons are considered over a finite space volume, defining a \textit{characteristic scale for nuclear pasta}.

One of the aims of this paper is to analyze the role of the Skyrme term when an external field is introduced. In the following section, this issue will be considered for both the exponential and Euler parametrization. In both cases, for $\lambda=0$ the equations of motion become unsolvable when the external field is very strong (tends to infinity). On the opposite, when $\lambda\neq0$, the equations of motion are always solvable. This fact is particularly evident in the Euler parameterization, which admits an ansatz that reduces the Skyrme equations to an ODE and keeps the layer's shape; this ansatz determines a universal solution, in the sense that it is independent of the external field (thus, it exists when the external field is zero, finite or infinite). The case of the baryonic tubes is quite different and harder to be solved (at least, analytically). It seems that the introduction of an external field does not allow for solutions that keep the baryonic tube structure, leading to incompatible sets of equations.


\section{Baryonic layers coupled to an external magnetic field} \label{LasagnaSection}


To emphasize the role of the Skyrme term in the stability of the Skyrmion, we are going to compare the NL$\sigma$M, which Lagrangian is defined by \eqref{ActionMaxwell} with $\lambda=0$, to the Skyrme model (with $\lambda\neq 0$). The Euler parameterization is very suitable for describing baryonic layers. Following \cite{NostroII}, we make the ansatz
\begin{align}
	U=e^{\Phi\kappa}e^{\chi\zeta}e^{\Theta\kappa},\label{ULasagna}
\end{align}
where, for now, $\chi$, $\Theta$, and $\Phi$ are general functions of space-time (a more specific ansatz on these functions will be defined in the following steps). $\kappa$ is an element of $\mathfrak{g}$ which can be written as
\begin{align}
	\kappa=\sum_{j=1}^r (c_j \lambda_j+c^*_j \tilde \lambda_j),
\end{align}
where $\lambda_j$ and $\tilde{\lambda}_j$ are eigenmatrices of the simple roots $\alpha_j$ and $-\alpha_j$ respectively. The quantities $c_j$ are complex numbers that can be chosen arbitrarily in a set of allowed values, which aim to guarantee the periodicity of the exponential of $\kappa$. The matrix $\zeta$ belongs to the Cartan subalgebra of $\mathfrak{g}$ (see \cite{Cacciatori:SUN,Cacciatori:EulerGen,NostroII}). With this choice of $U$, the left current becomes
\begin{align}\label{Lmu}
	\mathcal{L}_{\mu}&=e^{-\alpha \kappa}e^{-\xi \kappa}\Big[\partial_{\mu}%
	\alpha(\kappa-\hat{\kappa})\cr
	&\qquad\qquad\qquad+\partial_{\mu}\xi(\kappa+\hat{\kappa}%
	)+\partial_{\mu}\chi f\Big]e^{\xi \kappa}e^{\alpha \kappa},
\end{align}
where 
\begin{gather}\label{AlphaXi}
    \alpha=\frac{1}{2}(\Theta-\Phi),\quad\xi=\frac{1}{2}(\Theta+\Phi)
\end{gather}
and $\hat{\kappa}=e^{-\chi \zeta}\kappa e^{\chi \zeta}$. As shown in \cite{Canfora8:Lasagne,SergioFabrizio:2020zui,NostroII}, in the ungauged model the baryonic layers are obtained when the following orthogonality conditions are applied
\begin{align}\label{CondLasagne}
	\partial_\mu\Phi\partial^\mu\Theta=\partial_\mu\Phi\partial^\mu\chi=\partial_\mu\Theta\partial^\mu\chi=0.
\end{align}
If $\chi=\chi(r)$, the equations of motions (independently on the value of $\lambda$) reduce to one second order ODE (namely, $\chi(r)''=0$). 
In the presence of an external field, we expect that the ansatz \eqref{CondLasagne} has to be modified to get explicitly solvable equations. Furthermore, we cannot expect to be able to reduce the equations just to one ODE. The latter would be the optimal situation, but it could be incompatible with the presence of external fields. In what follows, we will see in the case of the Euler parameterization, the reduction of the equations of motion to one ODE is possible but it is not obvious if it is the same when the baryonic tubes are considered.


\subsection{Baryonic layers solutions in the NL$\sigma$M}\label{Sec:BaryionicLayer}


Let us consider the Lagrangian of the NL$\sigma$M
\begin{align}  \label{ActionMaxwellNLsM}
	L= \frac{K}{2}\mathrm{Tr}\left(\hat{%
		\mathcal{L}}_{\mu}\hat{\mathcal{L}}^{\mu}\right).
\end{align}
When the external field is zero and the conditions \eqref{CondLasagne} are applied, the equations of motions simplify to
{\begin{gather}
		\partial_{\mu}\partial^{\mu}\chi  -\sin(\chi)\left(\partial_\nu\alpha\partial^\nu\alpha-%
		\partial_\nu\xi\partial^\nu\xi\right)=0,\label{FirstSkyrmeNLL} \\
		\partial_{\mu}\partial^\mu\alpha=0\andd\partial_{\mu}\partial^\mu\xi=0.\label{SecondThirdSkyrmeNLL}
	\end{gather}
}
In \cite{NostroII}, the functions $\alpha$ and $\xi$ have been chosen as
{\begin{align}\label{AlphaXiLasagnaClassic1}
	\alpha&=\frac{q}{2}\theta-\frac{p}{2}\left(\frac{t}{L_\phi}-\phi\right),\\ \label{AlphaXiLasagnaClassic2}
    \xi&=\frac{q}{2}\theta+\frac{p}{2}\left(\frac{t}{L_\phi}-\phi\right),
\end{align}}
where $p$ and $q$ are integers. Moreover, $\chi=\chi(r)$. This way, \eqref{SecondThirdSkyrmeNLL} are automatically satisfied and \eqref{FirstSkyrmeNLL} reduces to $\chi''(r)=0$.

The external field can be introduced by using \eqref{HattedCurrent}, where we choose $T=\kappa$ as $U(1)$ gauge direction for the field $A_\mu$. This is equivalent to saying that the hatted quantities are obtained simply by replacing $\partial_\mu\alpha\rightarrow\partial_\mu\alpha-A_\mu$ in the ungauged terms. This remains true for the external field, with $A_\mu =A_\mu^{\mbox{ext}}$ (in Appendix \ref{AppGeneralEq} general equations for the Skyrme model have been obtained using the ansatz \eqref{ULasagna}. The analogous equations for the NL$\sigma$M are obtained by imposing $\lambda=0$).
Now, one may wish to conserve the advantages of the orthogonality conditions \eqref{CondLasagne} also in the gauged case. In this scope, we consider the following ansatz for the external field
\begin{align}\label{LasagnaACond}
	A_\mu^{\mbox{ext}}\partial^\mu\chi=0.
\end{align}
Hence, the gauged equations take the form
\begin{gather}
	\partial_{\mu}\partial^{\mu}\chi  -\sin(\chi)\left(\mathcal{A}_\nu\mathcal{A}^\nu-%
	\partial_\nu\xi\partial^\nu\xi\right)=0,  \label{FirstSkyrmeNLLA}\\
	\partial_{\mu}\mathcal{A}^\mu=0\andd\partial_{\mu}\partial^\mu\xi=0,
	\label{SecondThirdSkyrmeNLLA}
\end{gather}
where $\mathcal{A}_\mu=A_\mu^{\mbox{ext}}-\partial_{\mu}\alpha$. Notice that now the second term of \eqref{FirstSkyrmeNLLA} does not vanish when the choices \eqref{AlphaXiLasagnaClassic1} and \eqref{AlphaXiLasagnaClassic2} are applied, due to the presence of the external field.

The condition \eqref{LasagnaACond} and the first equation of \eqref{SecondThirdSkyrmeNLLA} are satisfied by the choice
\begin{align}\label{PotentialMagneticr(r)}
	A^{\mbox{ext}}_{\mu}=\left(0,0,A^{\mbox{ext}}_\theta(r),A^{\mbox{ext}}_\phi(r)\right)^T,
\end{align}
where $A^{\mbox{ext}}_\theta(r)$ and $A^{\mbox{ext}}_\phi(r)$ are linear functions of $r$, to give a constant external magnetic field, see \eqref{BConst}: \begin{align}
	\vec{B}^{\mbox{ext}}=\left(0,-\partial_rA^{\mbox{ext}}_\phi(r),\partial_rA^{\mbox{ext}}_\theta(r)\right)^T.
\end{align}
Let us define 
\begin{align}\label{PGen}
	P(t,r,\theta,\phi)=\mathcal{A}_\nu\mathcal{A}^\nu-%
	\partial_\nu\xi\partial^\nu\xi.
\end{align}
With our ansatz, $P(t,r,\theta,\phi)\equiv P(r)$, where
\begin{align}\label{Pansr}
	P(r)=A_\mu^{\mbox{ext}}A^{\mbox{ext},\mu}-\frac{q}{L_\theta^2} A^{\mbox{ext}}_\theta+\frac{p}{L_\phi^2} A^{\mbox{ext}}_\phi.
\end{align}
Thus, the system of equations \eqref{FirstSkyrmeNLLA} and \eqref{SecondThirdSkyrmeNLLA} becomes a second order ODE for the profile $\chi$,

\begin{align}
	\chi''(r)  -\sin(\chi)P(r)=0. \label{NLsMODE}
\end{align}

Notice that $P(r)$ is a second-order polynomial in $r$. \eqref{NLsMODE} does not allow us to define an analytical expression of the solution. Anyway, it is possible to define the expression for the energy density and the baryonic density, which shows that the shape of the baryonic layer is conserved. Moreover, the total baryonic charge can be computed exactly, since it only depends on the boundary conditions. 

\subsubsection{The energy density and baryonic charge}
The energy density is given by \eqref{TmunuSUN}. With our ansatz, it takes the form
	\begin{align}
		\rho_E&= \frac{K}{2}\|c\|^2\Big\{ \frac{8p^2}{L_\phi^2}+\frac{\chi'^2}{L_r^2}+\frac{q^2}{L_\theta^2}+4P(r)\sin^2\left(\frac{\chi}{2}\right)\Big\}.\label{vediamo}
	\end{align}
The baryonic density $\rho_B$ is given by \eqref{rhoB}, which becomes
\begin{align}
	\rho_B=-\frac{\|c\|^2}{4\pi^2}pq\ \partial_r%
	\cos(\chi). \label{rhoBNL}
\end{align}
As expected, it is a divergence, so its integral over the space variables depends only on the boundary conditions, which have been fully studied in \cite{NostroII}. 
An explicit computation leads to
\begin{align}\label{BGen}
	B
	=2I\eta{\|c\|^2},\quad\mbox{with}\quad I=pqn,
\end{align}
where $p$, $q$ and $n$ are integers.
This is the usual form for the baryonic charge in the ungauged case, which takes integer values (see \cite{NostroII}). 

\

\subsubsection{The large external magnetic field limit}
Let us look at the behavior of the ODE when the external field becomes very large. Let us call
\begin{align}
	A_\theta^{\mbox{ext}}=a_\theta r+b_\theta\andd A_\phi^{\mbox{ext}}=a_\phi r+b_\phi.
\end{align}
When the coefficients $a_\theta$ and $a_\phi$ becomes much larger than $1/r$, for nonzero $r$, the function \eqref{Pansr} tends to
\begin{align}
	P(r)\rightarrow P_l(r)=(a_\theta^2+a_\phi^2)r^2.
\end{align}
This way, equation \eqref{NLsMODE} loses coercivity. We will see in the next section that the introduction of the Skyrme term saves the situation; in particular, it is always possible to find an analytical solution for the Skyrme equations also for big values of the external field.


\subsection{Baryonic layers solutions in the Skyrme model}\label{LasagnaSol}


Let us consider the complete Lagrangian of the Skyrme model given in \eqref{ActionMaxwell}. The complete equations obtained for the ansatz \eqref{ULasagna} are reported in Appendix \ref{AppGeneralEq}. 

It is straightforward to show that the same choices defined in the previous section, summarized in the equations \eqref{AlphaXiLasagnaClassic1}, \eqref{AlphaXiLasagnaClassic2}, \eqref{PotentialMagneticr(r)} and $\chi=\chi(r)$, can be used to reduce the Skyrme equations to one ODE for the profile $\chi(r)$, which can be expressed as
	\begin{align}\label{FirstLasagnaSkyrme}
		&\chi'' \left\{1+\lambda\left[%
		P(r)+\frac{q^2}{4L_\theta^2}%
		\right]\right\}-\sin(\chi)\left(1-\frac{\lambda}{4}\chi'^2%
		\right)P(r)\cr
		&\quad-\lambda\sin(\chi)\cos(\chi)\Big\{\left(P(r)+\frac{q^2}{4L_\theta^2}\right)\frac{q^2}{4L_\theta^2}\cr
		&\qquad\quad-\Big[\frac{q}{2L_\theta^2}\left(A_\theta^{\mbox{ext}}-\frac{q}{2}\right)+\frac{p}{2L_\phi^2}A^{\mbox{ext}}_\phi
		\Big]^2\Big\} =0,
	\end{align}
where $P(r)$ is the same as in \eqref{Pansr}. 
This equation can be analytically solved if a further condition is imposed. Indeed, the last two lines of \eqref{FirstLasagnaSkyrme} are zero when 
\begin{align}\label{Condpq}
    {p}/{L_\phi^2}={q}/{L_\theta^2}. 
\end{align}
When this condition is applied, the first equation reduces to
{\small 
	\begin{align}\label{FirstLasagnaSkyrmeR}
		&\frac{\chi''}{L_r^2} \left\{1+\lambda\left[%
		P(r)+\frac{q^2}{4L_\theta^2}%
		\right]\right\}\cr & \qquad-\sin(\chi)\left(1-\frac{\lambda}{4L_r^2}\chi'^2\right)P(r) =0.
	\end{align}
}
The solution is given by
\begin{align}\label{FirstLinearSol}
	\chi(r)=\sqrt{\frac{4L_r^2}{\lambda}}r+\chi_0,
\end{align}
where $\chi_0$ is an integration constant. The boundary conditions necessary for the existence of the solition require that $\chi(0)=0$ and $\chi(2\pi)=\pi$. The first condition imposes that $\chi_0=0$; moreover, the second condition is satisfied when $\frac{\lambda}{L_r^2}=16$, which means that the constant of the model $\lambda$ and $L_r$ are not free. We are going to call \eqref{FirstLinearSol} as \textit{universal solution}, since it does not depends on the external field value.
 
Other solutions, different from \eqref{FirstLinearSol}, can be considered, which do not require that ${\lambda}/{L_r^2}= 16$; these solutions are piecewise-linear, with slope $\sqrt{{4L_r^2}/{\lambda}}$, and zero elsewhere. The only necessary condition in this case is that the slope of each linear part is bigger than the slope of the solution which does not present constant pieces. In the latter, we know that a necessary condition is ${\lambda}/{L_r^2}=16$; thus, in particular, the slope becomes $\sqrt{{1}/{4}}$. In the piecewise-linear solutions, the linear parts must undergo the condition $\sqrt{{4L_r^2}/{\lambda}}>\sqrt{{1}/{4}}$, which reads ${\lambda}/{L_r^2}<16$. 

It is worth noticing here that all these solutions exist only with $\lambda\neq 0$. Moreover, they are all independent of the value of the external field. This means that they always exist, even when the external field is zero or takes big values. As we are going to see in the next paragraph, this set of solutions defines baryonic layers; thus, it is remarkable the fact that the presence of the Skyrme term \textit{saves} the existence of baryonic layers when the external field is very big. This is not an obvious result; for instance, as we are going to observe, in the case of the baryonic tubes the situation becomes more complicated. 

In the following paragraphs, we are going to use this result to analyze the main characteristics of the gauged baryonic layers.

\subsubsection{The energy density and baryonic charge}
The energy density associated with the solution of the Skyrme equations is 
\begin{widetext}
	{\small 
		\begin{align}
			\rho_E=& \frac{K}{2}\|c\|^2\bigg\{ \frac{8p^2}{L_\phi^2}+\frac{q^2}{L_\theta^2}+\frac{\chi'^2}{L_r^2} +4P(r)\sin^2\left(\frac{\chi}{2}\right)\bigg\}\cr 
			&+\frac{K}{2}\|c\|^2\lambda\left(\frac{2p^2}{L_\phi^2}\right)\bigg\{%
			\left[\left(P(r)+\frac{q^2}{2L_\theta^2}\right)%
			-2%
			\mathcal{A}_\rho\partial^{\rho}\xi\right]\sin^2(\chi)+\frac{\chi'^2}{L_r^2}\bigg\}\cr 
			& +\frac{K}{2}\|c\|^2\lambda \bigg\{ \left[\left(P(r)+\frac{q^2}{4L_\theta^2}\right)\frac{q^2}{4L_\theta^2}-(\mathcal{A}_\rho\partial^\rho\xi)^2\right]\sin^2(\chi)  %
			+\left[P(r)\sin^2\left(\frac{%
				\chi}{2}\right)+\frac{q^2}{4L_\theta^2}\right]\frac{\chi'^2}{L_r^2}\bigg\}.
		\end{align}
	}
\end{widetext}
The baryonic charge remains of the form \eqref{rhoBNL}; the contribution brought by the Skyrme term is considered through the solution to the Skyrme equations. Nevertheless, the usual choice of the boundary conditions, defined by \eqref{Ranges}, leaves the shape of $B$ unchanged.

\subsubsection{The thermodynamics of the universal solution}
As discussed above, under certain conditions, it is always possible to find a universal solution to the Skyrme equations. Let us consider the case in which ${\lambda}/{L_r^2}=16$. Thus, the solution is represented by
\begin{align}
	\chi(r)=\frac{r}{2}.
\end{align}
Moreover, when the only non-zero component of the external potential is the $\theta$-component, $\Phi(u)$ is not necessarily a linear function of $u$. Let us define $A_\theta^{\mbox{ext}}=hr+h'$, where $h$ and $h'$ are arbitrary constants. This way,
\begin{align}
	\vec{B}^{\mbox{ext}}=(0,0,h)
\end{align} 
and
\begin{align}
	P(r)
	&=\frac{1}{L_\theta^2}\left[h^2 r^2+h\left(2h'-q\right)r+h'(h'-q)\right]
\end{align}

\begin{figure}[htbp!]
		\hspace{-1pt}\includegraphics[scale=.7]{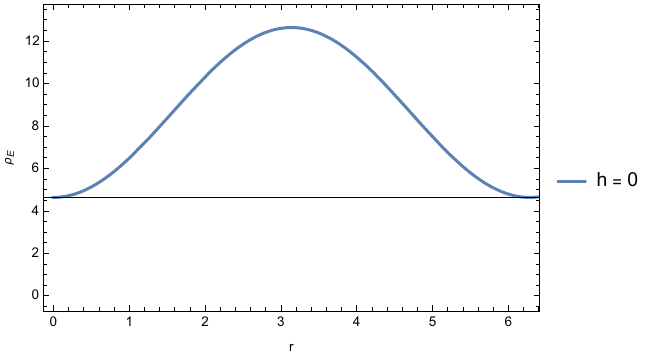}\\
        \hspace{7pt}\includegraphics[scale=.7]{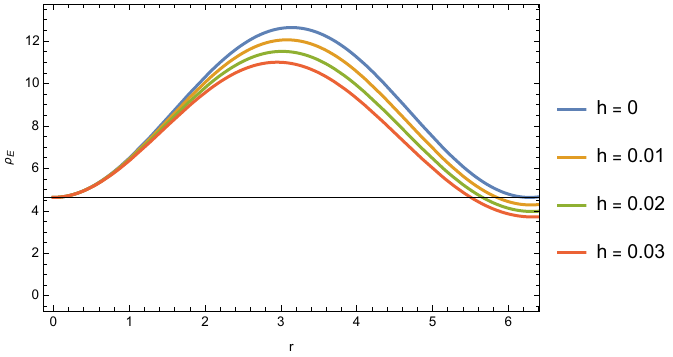}
		\caption{Here is shown the energy density of the universal solution of baryonic layers for different values of $h$ when $0\leq r \leq 2\pi$. The upper Figure represents the \textit{uncoupled} case, with $h=0$ In the lower Figure, the uncoupled solution is compared to the other cases with $h\neq 0$. The parameters have been set to the following values: $L_r=L_\theta=L_\phi=1$, $K=2$, $\|c\|^2=1$, and $p=q=1$.}
		\label{Fig:LinearLasagnaDifferntA}
	\end{figure}

Collecting the terms relative to the external field contribution and imposing that $h'=0$ for simplicity, the energy density takes the form 
\begin{widetext}
	{\small 
		\begin{align}\label{rhoEdipA}
			\rho_E=& \rho_0+hr\rho_1+h^2r^2\rho_2,
		\end{align}
	}
	where
	{\small 
		\begin{align}
			\rho_0&=K\|c\|^2 \bigg\{\frac{p^2}{2L_\phi^2}\left[1+\frac{\lambda}{8L_r^2}+\frac{q^2\lambda}{2L_\theta^2}\sin^2\left(\frac{r}{2}\right)\right]+\frac{q^2}{4L_\theta^2}\left(1+\frac{\lambda}{8L_r^2}\right)+\frac{1}{8L_r^2}\bigg\},\\
			\rho_1&= -K\|c\|^2\bigg\{\frac{p^2}{2L_\phi^2}\left[\frac{q\lambda}{L_\theta^2}\sin^2\left(\frac{r}{2}\right)\right]+\frac{q}{L_\theta^2}\sin^2\left(\frac{r}{4}\right)\left(1+\frac{\lambda}{8L_r^2}\right)\bigg\},\\
			\rho_2&=K\|c\|^2\bigg\{\frac{p^2}{2L_\phi^2}\bigg[
			\frac{\lambda}{2L_\theta^2}\sin^2\left(\frac{r}{2}\right)\bigg]+\frac{1}{L_\theta^2}\sin^2\left(\frac{r}{4}\right)\left(1+\frac{\lambda}{8L_r^2}\right)\bigg\}.
		\end{align}
	}
\end{widetext}
It is worth noticing that the presence of the terms proportional to $\sin^2\left(\frac{r}{4}\right)$ in $\rho_1$ and $\rho_2$ change the periodicity of the energy density, which, in particular, becomes $4\pi$-periodic. 
This situation is represented in Figure \ref{Fig:LinearLasagnaDifferntA}, where the energy density at different values of $h$ is depicted for $0\leq r \leq 2\pi$. It appears evident that the energy density loses its periodicity. However, for small values of $h$ it conserves approximately the shape of the layer. For this reason, we are going to consider this displacement from the periodic solution as a perturbation. In future work, we are going to analyze the impact of the back-reaction on this behavior of the energy density: it is expected that the skyrmion tends to conserve its shape and contrast the external field, maintaining the periodicity.
The dependence of the total energy on the external field can be formulated by integrating $\rho_E$ over the volume of the box. Namely, 
\begin{align}
	E=L_rL_\theta L_\phi\int\rho_E drd\theta d\phi.
\end{align}
An explicit computation leads to the following form for the total energy, where we used the condition $p=q$ imposed by Eq. \eqref{Condpq}
\begin{align}\label{TEn}
	E(q;h)=64KL\|c\|^2\pi^2\sum_{i=1}^{4}\left[\Sigma_i(h)\right]q^i,
\end{align}
where
\begin{align}
	&\Sigma_0(h)=\frac{\pi}{4L_r^2}+h^2\frac{8\pi}{L_\theta^2}(\pi^2+6),\label{I} \\ 
	&\Sigma_1(h)=-h\bigg[\frac{6}{L_\theta^2}(4+\pi^2)\bigg],\label{II} \\
	&\Sigma_2(h)=\frac{3\pi}{L_\theta^2}+\frac{6\pi}{L_\phi^2}\pi+h^2\frac{16L_r^2\pi}{L_\phi^2L_\theta^2}\frac{2\pi^2-3}{3},\label{III} \\
	&\Sigma_3(h)=-h\frac{16L_r^2\pi^2}{L_\phi^2L_\theta^2},\label{IV} \\
	&\Sigma_4(h)=\frac{8L_r^2}{L_\phi^2L_\theta^2}\pi \label{V}
\end{align}
and $(q;h)$ indicates the dependence on the parameters $q$ and on the coefficient of the gauge field $h$ (the relation $\frac{\lambda}{16L_r^2}=1$ has been used). 

The baryonic charge is obtained by integrating the baryonic density
\begin{align}\label{TB}
	\rho_B=\frac{\|c\|^2}{8\pi^2}pq\ %
	\sin\left(\frac{r}{2}\right),
\end{align}
which gives
\begin{align}\label{LasagnaBaryonicCharge}
	B(p,q)=\frac{\|c\|^2}{2}pq
\end{align}

In the most general case, the expression for the partition function can be written as follows
\begin{align}\label{PartitionFunction}
	\mathcal{Z}=\sum_{p,q}\exp\left\{-\beta\left[ E(p,q;h)-\mu_BB(p,q)\right]\right\},
\end{align}
where $\beta={1}/{K_B T}$ ($T$ rbeing the temperature) and $\mu_B$ is the chemical baryonic potential. 

By assuming $L_\theta=L_\phi=L$ and the usual condition $p=q$, and using the expression \eqref{TEn} and \eqref{TB}, the partition function thus becomes
{\small\begin{align}
	&\mathcal{Z}_q=\exp\bigg\{-\beta\|c\|^2\cr
    &\qquad\qquad\quad\times\bigg[2KL_rL^2\pi^2\sum_{i=1}^{4}\Sigma_i(h)q^i-\frac{q^2}{2}\mu_B\bigg]\bigg\}.
\end{align}}
\vspace{2pt}

\subsubsection{The dependence on the external field}
\begin{figure*}\label{Fig:CausticRegions}
    \centering
	\includegraphics[scale=1]{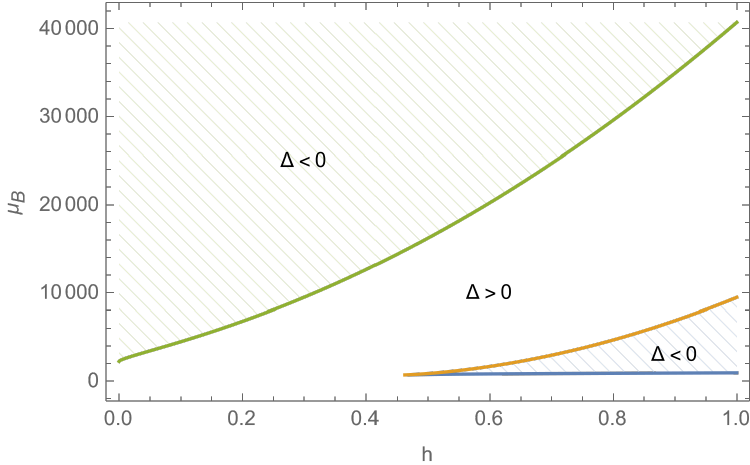}
	\caption{Here we show the regions separated by the caustic. The lines represent the branches of the caustic and the shaded parts are the regions in which $\Delta$ is negative. The parameters have been set to the following values: $L_r=L_\theta=L_\phi=1$, $K=2$, $\|c\|^2=1$, and $m=1$. 
 }
	\label{Fig:muBVSB}
\end{figure*}

We want to analyze in detail the behavior of the partition function in terms of the external field. We can write $\mathcal{Z}$ in the form 
\begin{widetext}
	\begin{align}
		\mathcal{Z}&=\sum_{q=-\infty}^{+\infty}\exp\bigg\{-\frac{\beta}{\eta_4^3}\bigg[\left(\eta_4q+\frac{\eta_3}{4}\right)^4+\left(\tilde{\eta}_2\eta_4-\frac{3}{8}\eta_3^2\right)\eta_4^2q^2\cr
		&\qquad\qquad\qquad\qquad\qquad\qquad\qquad\qquad\qquad\qquad+\left(\eta_1\eta_4^2-\frac{1}{16}\eta_3^3\right)\eta_4q+\eta_0\eta_4^3-\frac{\eta_3^4}{256}\bigg]\bigg\},
	\end{align}
where
	\begin{align}
	&\eta_i=2\|c\|^2KL_rL^2\pi^2\ \Sigma_i(h),\quad\mbox{for}\ i=0,1,2,3,4, \label{eta}\\
	&\tilde{\eta}_2=\|c\|^2\left[2KL_rL^2\pi^2\ \Sigma_2(h)-\frac{\mu_B}{2}\right]=\eta_2-\frac{\|c\|^2\mu_B}{2}. \label{tildeeta}
	\end{align}
\end{widetext}
Strictly speaking, our model is valid at low energies and, therefore, for small values of the temperature, aka large $\beta$. In this situation, for fixed values of all parameters, there are just one or a few dominating terms of the series and all others are exponentially smaller. One has simply to look for the absolute minimum of the quartic exponent as a function of the discrete variable $q$. On the other side, we need to consider such exponent as a function of $h$ and $\mu_B$, which makes our aim much harder to accomplish. Nevertheless, we can observe that considering instead small values of $\beta$ would allow us to replace the sum over $q$ with an integral, simpler to analyze from this viewpoint. At least for large values of $h$ and/or $\mu_B$, as we will see, this can be analyzed using the usual saddle point approximation, more or less, the same saddles we have just mentioned for the low-temperature case, at least for values of energies such that the solitonic solution still survives. For these reasons, our strategy will be to pass to the description in the limit of the continuum, and only at the end we will comment on which changes we do expect at low energies.\\
After the substitution
\begin{align}\label{IntegralAppox}
	\eta_4q\longrightarrow q'-\frac{\eta_3}{4},
\end{align}
the partition function at low temperatures is therefore described by the integral
\begin{align}\label{Z}
	\mathcal{Z}&=\int_{-\infty}^{+\infty}\frac{dq'}{\eta_4}\exp\bigg\{-a\left[q'^4+bq'^2+cq'+d\right]\bigg\},
\end{align}
where
\begin{gather}
	a=\frac{\beta}{\eta_4^3},\quad b=\tilde{\eta}_2\eta_4-\frac{3}{8}\eta_3^2,\cr  c=\eta_1\eta_4^2+\frac{\eta_3^3}{8}-\frac{\tilde\eta_2\eta_4\eta_3}{2},\cr d=\eta_0\eta_4^3-\frac{3\eta_3^4}{256}+\frac{\tilde\eta_2\eta_4\eta_3^3}{16}-\frac {\eta_1 \eta_4^2 \eta_3}4.
\end{gather}
Since the saddle points of the exponent in the integrand are relevant both in the discrete and in the continuum limit, before a more systematic analysis of this integral, let us first tell more about the stationary points. We expect that different values of the parameters $b$ and $c$ can determine regions of different phases of the baryonic layers. On the other hand, since all parameters are real, the contribution of the saddle points is influenced by their real or complex nature in general. They are solutions of the cubic equation
\begin{align}
    4z^3+2bz+c=0,
\end{align}
which has discriminant $\Delta=c^2+(2b/3)^3$. For $\Delta<0$ there are 3 real solutions, while for $\Delta>0$ there are one real and 2 complex conjugate solutions. 
$\Delta=0$ corresponds to the coalescence of at least two saddle points. It thus determines the \textit{caustic curve} 
\begin{align}\label{Caustic}
    \frac{b^3}{27}+\frac{c^2}{8}=0,
\end{align} 
which separates different phases of the baryonic layer. Indeed, since both $b$ and $c$ depend linearly on $\mu_B$ with $h$-dependent coefficient, \eqref{Caustic}
determines a relation between the chemical potential and the magnetic field, given by a cubic equation of the form
\begin{align}\label{CubicDiscriminant}
    0 = c_1(h)[\mu_B^3+c_2(h)\mu_B^2+c_3(h)\mu_B+c_4(h)],
\end{align}
for given functions $c_j(h)$. The detailed expression for this cubic can be found in App. \ref{app:hmu}. Notice that for $b>0$ $\Delta$ is always positive so the caustic must belong in the $b\leq0$ region. Moreover, in the equation, only $h^2$ appears explicitly, so we can restrict to consider only positive values of $h$.
The exact shape of the caustic curve thus depends on the existence of real roots of the cubic, therefore, from the sign of its discriminant.
In App. \ref{app:hmu}, it is shown that the discriminant is positive for $0\leq h\leq \bar h$, with
\begin{align}
    \bar h=4 \frac {\eta_4}{\hat\eta_3} \sqrt{\frac {\hat\eta_1}{\hat\eta_3}},
\end{align}
where $\hat \eta_j=\eta_j/h$, $j=1,3$, are constants. Therefore, for $h<\bar h$ it appears only one branch of the caustic, while a second double upper branch appears for $h\geq \bar h$, as represented in Figure \ref{Fig:CausticRegions}. 

In each portion of this phase space \eqref{Z} may have different asymptotic behaviors when $b$ and $c$ are running toward infinity in different directions, possibly manifesting the \textit{Stokes phenomenon}. Let us briefly recall that from a mathematical point of view the Stokes phenomenon appears each time one is solving a
differential equation (system) presenting at least one irregular singular point. The asymptotic behavior of any solution near the singular point has not a definite
behavior but depends on different angular sectors approaching the point, separated by the so-called Stokes walls. A very simple explanation of this phenomenon can be
found in \cite{SIAM}. The prototype case of such phenomenon is given by the Airy function that, despite being an entire function, presents a complicated asymptotic
behavior at infinity, \footnote{Indeed, the phenomenon was discovered by G.G. Stokes in \cite{Stokes1}, \cite{Stokes2} by studying the Airy integrals arising in the computation of the intensity of light near a caustic, \cite{AiryCaustic}.}. To fix our conventions, let us recall that what happens is that at infinity the Airy function takes the form $A_i(z)=c_1 u_1(z)+c_2 u_2(z)$, where $c_j$ are constants, while $u_j$ are polydromic functions. Thus, the $u_j$ are separately defined on a cut plane and transformed according to a nontrivial monodromy after a change of phase of $2\pi$ around infinity. However, $Ai$ is monodromic, hence, after such phase rotation, it must be described by a new combination of the $u_j$, to preserve the monodromy. Therefore, the constant $c_j$ has to change (discontinuously, being constants) under such an operation. This is essentially the Stokes phenomenon. During the phase rotation, one crosses lines that determine the main change of the asymptotic behavior of the function: along certain lines one has a maximal monotonic decrease or increase of the modulus of the functions (essentially when one of the two functions dominates), along other lines one has an oscillating behavior (where the two functions have comparable modulus). We call the former {\em the anti-Stokes lines} and the latter {\em the Stokes lines}. If $u_j\sim e^{\phi_j}$, then anti-Stokes lines are defined by $Im(\phi_1-\phi_2)=0$, while the Stokes lines are defined by $Re(\phi_1-\phi_2)=0$. In literature, one easily meets both these and the opposite convention (where the notion of Stokes and anti-Stokes are interchanged). 
In practice, the (anti-)Stokes lines represent walls, crossing which different saddle points enter the game in determining the asymptotic behavior in a saddle point approximation. This is largely studied in the mathematical literature,
whenever one has to tackle the asymptotic expansion of solutions of differential systems \cite{holmes} \cite{eastham} \cite{olver} \cite{wong}, and it has several applications. For example, it appears as a tool in Topological Quantum Field Theories (see e.g. \cite{Gu:2023bjg}, \cite{TQFT} and references therein), in mirror symmetry and related topics \cite{Kontsevich:2013rda}, in Chern-Simons theory \cite{Witten:2010cx}, in Bridgeland stability \cite{Bridgeland_2011}, in Quantum Cohomology \cite{Guzzetti}, etc.\\
A particular field of application of interest for the present application is resurgence theory \cite{kowalenko}. In this case, indeed, one uses Borel's resummation methods to deal with diverging infinite series, then determining quantitative relations between perturbative and non-perturbative data. Since saddle points typically catch non-perturbative information, here is where the Stokes phenomenon plays a major role. In this way, Stokes lines provide a contact between perturbative and non-perturbative regimes and we have to expect that they do not change sensitively in passing from one regime to the other. Since Skyrme theory is a low-energy effective description of QCD, we thus expect that analyzing the Stokes phenomenon in the Skyrme regime gives interesting information in the full QCD regime and vice versa.

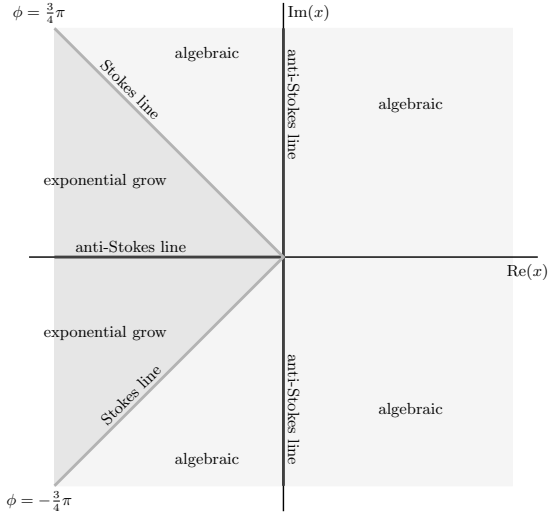
\begin{figure}[!htbp]
\begin{center}
\resizebox{7.5cm}{!}{
\begin{tikzpicture}[>=latex]  
\filldraw [gray!20!white] (-4.5,4.5)--(0,0)--(-4.5,-4.5)--cycle;
\filldraw [gray!8!white] (-4.5,4.5)--(0,0)--(-4.5,-4.5)--(4.5,-4.5)--(4.5,4.5)--cycle;
\draw [thick] (-5,0)--(5,0);
\draw [thick] (0,-5)--(0,5);
\draw [ultra thick,gray!50!black] (0,-4.5)--(0,4.5);
\draw [ultra thick,gray!50!black] (-4.5,0)--(0,0);
\draw [ultra thick,DarkSand] (-4.5,4.5)--(0,0)--(-4.5,-4.5);
\node at (0.5,4.8) {${\rm Im}(x)$};
\node at (4.8,-0.3) {${\rm Re}(x)$};
\node at (-4.8,4.8) {$\phi=\frac 34 \pi$};
\node at (-4.8,-4.8) {$\phi=-\frac 34 \pi$};
\node at (-3,0.2) {anti-Stokes line};
\node [rotate=-90] at (0.2,3) {anti-Stokes line};
\node [rotate=-90] at (0.2,-3) {anti-Stokes line};
\node [rotate=-45] at (-3,3.3) {Stokes line};
\node [rotate=45] at (-3,-2.7) {Stokes line};
\node at (-3.5,1.5) {exponential grow};
\node at (-3.5,-1.5) {exponential grow};
\node at (2.5,3) {algebraic};
\node at (2.5,-3) {algebraic};
\node at (-1.5,4) {algebraic};
\node at (-1.5,-4) {algebraic};
\end{tikzpicture}
}
\end{center}
\caption{Stokes' lines and anti-Stokes' lines in the complex $x$-plane. Crossing the Stokes lines the function passes from an algebraic/oscillating behavior to an exponentially growing regime. In restricting to real variables we see that negative values exactly belong to an anti-Stokes line (maximally growing).
Here $\phi=\arg(x)$}\label{xinfinito}
\end{figure}

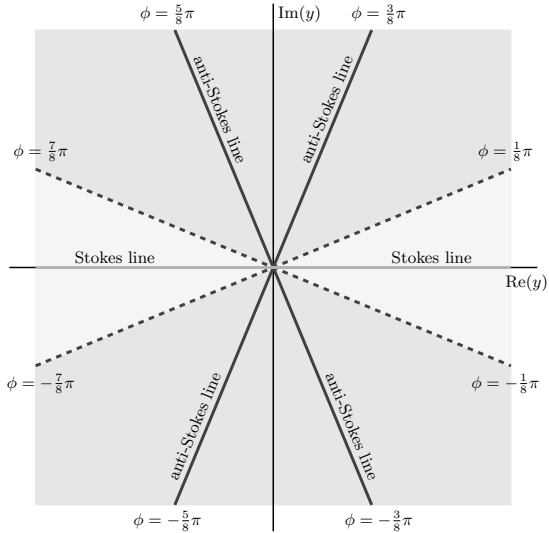
\begin{figure}[!htbp]
\begin{center}
\resizebox{7.5cm}{!}{
\begin{tikzpicture}[>=latex]  
\filldraw [gray!20!white] (0,0)--(4.5,4.5*0.414213562)--(4.5,4.5)--(-4.5,4.5)--(-4.5,4.5*0.414213562)--cycle;
\filldraw [gray!20!white] (0,0)--(4.5,-4.5*0.414213562)--(4.5,-4.5)--(-4.5,-4.5)--(-4.5,-4.5*0.414213562)--cycle;
\filldraw [gray!8!white] (0,0)--(4.5,4.5*0.414213562)--(4.5,-4.5*0.414213562)--cycle;
\filldraw [gray!8!white] (0,0)--(-4.5,4.5*0.414213562)--(-4.5,-4.5*0.414213562)--cycle;
\draw [thick] (-5,0)--(5,0);
\draw [thick] (0,-5)--(0,5);
\draw [ultra thick,gray!50!black] (-4.5*0.414213562,-4.5)--(4.5*0.414213562,4.5);
\draw [ultra thick,gray!50!black] (-4.5*0.414213562,4.5)--(4.5*0.414213562,-4.5);
\draw [ultra thick,gray!50!black,dashed] (-4.5,-4.5*0.414213562)--(4.5,4.5*0.414213562);
\draw [ultra thick,gray!50!black,dashed] (-4.5,4.5*0.414213562)--(4.5,-4.5*0.414213562);
\draw [ultra thick,DarkSand] (-4.5,0)--(4.5,0);
\node at (0.5,4.8) {${\rm Im}(y)$};
\node at (4.8,-0.3) {${\rm Re}(y)$};
\node at (2,4.8) {$\phi=\frac 38 \pi$};
\node at (-2,4.8) {$\phi=\frac 58 \pi$};
\node at (4.4,2.2) {$\phi=\frac 18 \pi$};
\node at (-4.4,2.2) {$\phi=\frac 78 \pi$};
\node at (2,-4.8) {$\phi=-\frac 38 \pi$};
\node at (-2,-4.8) {$\phi=-\frac 58 \pi$};
\node at (4.4,-2.2) {$\phi=-\frac 18 \pi$};
\node at (-4.4,-2.2) {$\phi=-\frac 78 \pi$};
\node at (3,0.2) {Stokes line};
\node at (-3,0.2) {Stokes line};
\node [rotate=67.5] at (1,3) {anti-Stokes line};
\node [rotate=-67.5] at (1.5,-3) {anti-Stokes line};
\node [rotate=-67.5] at (-1,3) {anti-Stokes line};
\node [rotate=67.5] at (-1.5,-3) {anti-Stokes line};
\end{tikzpicture}
}
\end{center}
\caption{Stokes' lines and anti-Stokes' lines in the complex $y$-plane. The dark sectors correspond to exponential growth, while the light sector to exponential decay. The dashed lines are not Stokes lines but transition lines in the sense of Poincar\'e, see \cite{Paris91}. Here $\phi=\arg(y)$. Notice that in in our specific case $y=ia^{\frac{3}{4}}c$ is purely imaginary, so in the exponential growing case, and, because of the symmetry, there is no trace of the Stokes transition in passing from positive to negative values.}\label{yinfinito}
\end{figure}

To analyze our specific case, it is convenient to consider the change of variable $q=a^{\frac 14}q'$. In this way, \eqref{Z} takes the form
\begin{align}
    \mathcal{Z}&=\frac {e^{-i\frac \pi8}}{\eta_4 a^{\frac 14}} \mathcal{P}(x,y),
\end{align}
where
\begin{align}
	\mathcal{P}(x,y)=e^{i\frac \pi8}\int_{-\infty}^{+\infty}dq\exp\bigg\{-q^4-xq^2+iyq\bigg\}, \label{PIntegral}
\end{align}
is the Pearcey's integral, as redefined in \cite{Paris91}, with
\begin{align}
	x=a^{\frac{1}{2}}b\andd y=ia^{\frac{3}{4}}c.
\end{align}
As we already mentioned, we are working for small values values of $a$. Therefore, in general, we expect $x$ and $y$ to be small. Nevertheless, nothing prevents us from considering high values of $h$ and/or $\mu_B$ enough to still have large values of $x$ and $y$. Since these are exactly the cases where the Stokes phenomenon is expected to arise, here we consider exactly such a situation. 
Since the Pearcey integral has been largely studied, we can easily analyze its behavior in the regions of our interest by referring to the suitable literature,
see App. \ref{App:pearcey}.\\[0.4cm]
\textbf{Asymptotic expansion for \boldmath{$|x|\rightarrow\infty$}.} 
When $|x|\to\infty$  with bounded $y$, one gets a uniform qualitative behavior in the complex $x$ plane, in the sense that Stokes lines do not depend on $y$.
A representation is provided in Fig. \ref{xinfinito}.\\[0.4cm]
\textbf{Asymptotic expansion for \boldmath{$|y|\rightarrow\infty$}.} 
When $|y|\to\infty$  with bounded $x$, we have an almost uniform qualitative behavior, now in the complex $y$ plane. Recall that the Pearcey integral is an even 
function of $y$.  
A representation of the Stokes lines is provided in Fig. \ref{yinfinito}.

\textbf{Asymptotic behavior around the caustic curve.}
In this case, we get that near the cuspid, for large $x$, it is dominating an exponentially growing mode. Aside from it, two other modes, decaying exponentially at infinity,
have an oscillating character where $\Delta>0$, while are algebraic in $\Delta<0$. This behavior is quite different than the one in \cite{Kaminski}, where he gets
instead three oscillating terms with decay $x^{-\frac 12}$ for $\alpha>0$ and one oscillating term and two exponentially decaying for $\alpha<0$. This is the behaviour
at $x=|x|e^{-i\frac \pi4}$ in our coordinates. Notice that it is exactly this difference in the phase that makes the difference. Indeed, the always oscillating mode
in \cite{Kaminski} is the one corresponding to the critical point $z_1$, and it is oscillating only for the above phase. As soon as $x=|x|e^{-i\phi}$ with 
$0<\phi<\frac \pi4$ (which is still in the region of the definition of their expansion as well as ours), the two results agree, and this mode becomes exponentially growing.\\[0.4cm]
\textbf{An important remark.} In our setting we considered the extension of the $x$ and $y$ variables to complex values. This led us to meet the Stokes phenomenon, 
which, however, appears to be related to physical situations only when we restrict ourselves to real values. However, our model is understood to describe low-energy
QCD at finite temperatures. It is a known fact that in a full non-perturbative setting, that is in lattice computation, a real chemical potential is problematic, and
the solution consists of replacing it with a pure imaginary chemical potential \cite{AKW}. More in general, in \cite{Kerbstein} it has been shown that from the partition
function with a purely imaginary chemical potential one can get a complete picture of the phase space. In \cite{Roberge-Weiss}, the same technique combined with numerical computation has been used to get a complete description of the Roberger-Weiss transition. Therefore, we conclude that the above discussions of the Pearcey integral remain physically valid for generically complex $x$ and $y$ parameters since they just correspond to an imaginary chemical potential.
\subsubsection{Free energy and susceptibility}\label{sec:suscept}
The previous analysis of the partition functions allowed us to study the behaviors of the physical properties of the skyrmionic layers. In particular, in this section, we are going to analyze the free energy $F(h)$ and the susceptibility $\chi(h)$ in terms of the external field $h$, the temperature $\frac{1}{\beta}$, and the magnetic moment $\mu$.

\begin{figure*}[!htbp]
    \centering
	\includegraphics[scale=1]{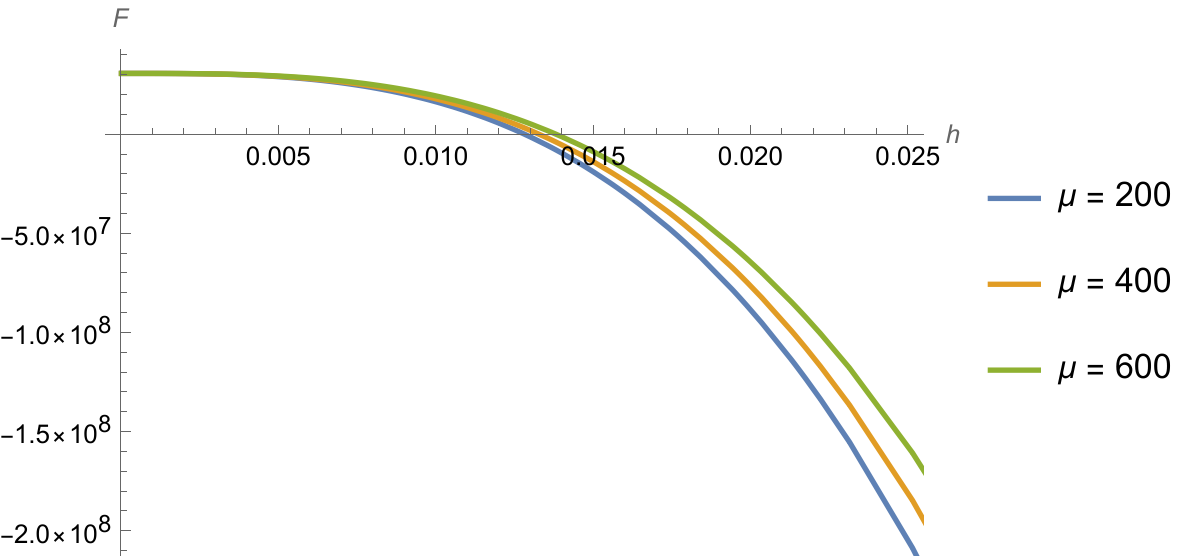}
	\caption{In this picture is represented the Free Energy in terms of the external field for different values of the magnetic moment $\mu$. Here, $\beta=10^{-4}$. It can be noticed that for certain values of the external field $h$, the free energy takes negative values.}\label{Fig:FMu}
\end{figure*}

In Fig. \ref{Fig:FMu}, different plots of the free energy in terms of the external field $h$ and for different values of the magnetic moment $\mu$ are represented. Here, it is important to observe that the function $F(h)$ takes negative values when $h$ becomes big. It is necessary to point out that our model must be considered in the range of small values of the external field since we are neglecting the internal back-reaction of the skyrmion, which will be the subject of future works.

\begin{figure*}[!htbp]
    \centering
	\includegraphics[scale=1]{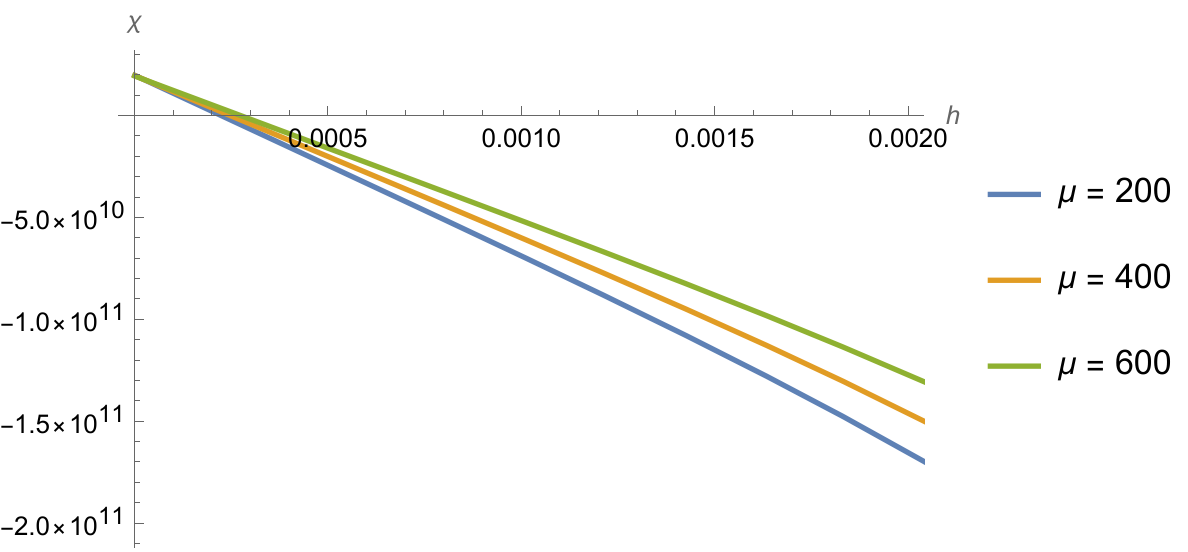}
	\caption{In this picture is represented the susceptibility in terms of the external field for different values of the magnetic moment $\mu$. Here, $\beta=10^{-4}$. For small values of $h$, it takes positive values and becomes negative after a certain limit.}\label{Fig:ChiMu}
\end{figure*}

In Fig. \ref{Fig:ChiMu}, it is depicted the behavior of the susceptibility in terms of the external field for $\beta=10^{-4}$. It is evident that $\chi$ takes positive values for small $h$. It can be due to different reactions of the skyrmion at different values of the external field, but it is worth remarking, as discussed in the perspectives, that we have not yet fully included the back reaction here.

\section{Chiral conformal field theory of baryonic layers from the Skyrme model}


In \cite{Canfora15:CFT}, the authors introduce an extension of the ansatz \eqref{AlphaXiLasagnaClassic1} and \eqref{AlphaXiLasagnaClassic2}, which consists in a generalization of the function $\Phi(t,\phi)$. Indeed, let us observe that the orthogonality conditions \eqref{CondLasagne} are preserved when $\Phi(u)$ is a general function of $u=t/L_{\phi}-\phi$. To keep track of the contributions brought by this generalization, we write
\begin{align}
	\Phi(u) = \tilde\Phi(u)+pu,
\end{align}
where $\tilde\Phi(u)$ corresponds to the general function of $u$ and the last term has the usual form of the \textit{old} $\Phi$. The introduction of the new term $\tilde\Phi(u)$ leads to the definition of a chiral conformal field theory (CCFT), as stated in \cite{Canfora15:CFT}. In this section, we want to study the contribution of the conformal part to the partition function under the action of the eternal field (in \cite{Canfora15:CFT}, this possibility has been analyzed in the context without an external field).

In section \ref{LasagnaSol}, we defined some conditions that lead to a significant simplification of the Skyrme equations. In particular, we considered the condition \eqref{Condpq}. When the conformal term is added, a stronger condition is necessary. Indeed, the polynomial $P(r)$ that appears in \eqref{FirstLasagnaSkyrme} (and that have been introduced in \eqref{PGen}) takes a more general form
\begin{align}
	P(t,r,\phi)=A_\mu^{\mbox{ext}}A^{\mbox{ext},\mu}-2\partial_\mu\alpha A^{\mbox{ext},\mu},
\end{align}
where the dependence on $t$ and $\phi$ is enclosed in $\Phi(u)$, contained in $\alpha$, and the dependence on r is due to $A_\mu^{\mbox{ext}}$. Differently from the other case (with $\tilde{\Phi}=0$), the dependence on $t$ and $\phi$ does not cancel. Therefore, also when ${p}/{L_\phi^2}={q}/{L_\theta^2}$, the last term of \eqref{FirstLasagnaSkyrme} does not cancel. To re-conduce the Skyrme equation to the simplified form \eqref{FirstLasagnaSkyrmeR}, we need the further condition $A_\phi^{\mbox{ext}}=0$. This way,
\begin{align}
	P(t,r,\phi)\rightarrow P(r)=\frac{{A_\theta^{\mbox{ext}}}}{L_\theta^2}\left(A_\theta^{\mbox{ext}}-q\right),
\end{align} 
and the Skyrme equations are simplified to 
{\small 
	\begin{align}
		&\frac{\chi''}{L_r^2} \left\{1+\lambda\left[%
		P(r)+\frac{q^2}{4L_\theta^2}%
		\right]\right\}\cr & \qquad-\sin(\chi)\left(1-\frac{\lambda}{4L_r^2}\chi'^2\right)P(r) =0.
\end{align}}
Notice that also this case we can find the usual universal solution analyzed in the previous section.

Let us discuss how the baryonic charge, the energy density, and the partition function generalize to the conformal case.

First of all, it is straightforward to observe that the conformal term does not contribute to the \textit{baryonic charge}. Indeed, the baryonic density can be written as
\begin{widetext}
\begin{align}
	\rho_B^{CCFT}= \rho_B+
		\frac{\|c\|^2}{8\pi^2}q\partial_\phi\tilde{\Phi}%
		\sin\left(\frac{r}{2}\right)
\end{align}
\end{widetext}
where $CCFT$ labels the \textit{conformal baryonic density} and $\rho_B$ has been defined in \eqref{rhoBSm}. We can consider some periodic conditions over the function $\tilde{\Phi}$. Namely,
\begin{gather}
	\tilde{\Phi}(t,\phi=0)=\tilde{\Phi}(t,\phi=2\pi),\\
	\partial_\phi\tilde{\Phi}(t,\phi=0)=\partial_\phi\tilde{\Phi}(t,\phi=2\pi).
\end{gather}
This way, the integral over $\phi$ of $\tilde{\Phi}$ cancels.

Let us discuss the generalization of the \textit{energy density}. It takes the usual general form defined by Eq. \eqref{rhoEdipA}, but with
\begin{widetext}
	{\small 
		\begin{align}
			\rho_0&=K\|c\|^2 \bigg\{\mathcal{I}\left[1+\frac{\lambda}{8L_r^2}+\frac{q^2\lambda}{2L_\theta^2}\sin^2\left(\frac{r}{2}\right)\right]+\frac{q^2}{4L_\theta^2}\left(1+\frac{\lambda }{4L_r^2}\right)+\frac{1}{16L_r^2}\bigg\},\\
		\end{align}
	}
    where
	\begin{align}
		\mathcal{I}=\frac{1}{4}\left((\partial_t\tilde{\Phi})^2+\partial_i\tilde{\Phi}\partial^i\tilde{\Phi}\right)+\partial_t\tilde{\Phi}\frac{p}{L_\phi}+\frac{p^2}{2L_\phi^2}.
	\end{align} 
\end{widetext}
Thus, the function $\tilde{\Phi}(u)$ contributes only inside the factor $\mathcal{I}$. The integration of the energy density over the volume leads to the following form of the total energy
\begin{align}
	E(q;h,m)&=2\pi^2KL_rL_\theta L_\phi\|c\|^2\cr
	&\quad\cdot\sum_{i=1}^{4}\left[e_0\tilde{\Sigma}_i(h)+\Sigma_i(h)\right]q^i,
\end{align}
where $\Sigma_i(h)$ take the same form as in \cref{I,II,III,IV,V} and
{\small
	\begin{align}
        &\tilde{\Sigma}_0(h)=3\pi
		+h^2\frac{8L_r^2\pi}{L_\theta^2}\left[\frac{2\pi^2-3}{3}\right],\\
        &\tilde{\Sigma}_1(h)=-h\frac{8L_r^2\pi^2}{L_\theta^2}\\
        &\tilde{\Sigma}_2(h)=\frac{4L_r^2}{L_\theta^2}{\pi},\\
        &\tilde{\Sigma}_3(h)=\tilde{\Sigma}_4(h)=0.
\end{align}}
Moreover,
\begin{align}
	e_0=\frac{1}{2\pi}\int_{0}^{2\pi}\partial_{t}\tilde{\Phi}^2+\partial_{i}\tilde{\Phi}\partial^{i}\tilde{\Phi}\ d\phi.
\end{align}
Notice that $\tilde{\Phi}(u)=0$ implies $e_0=0$. As already proposed in \cite{Canfora15:CFT}, the terms $e_0$ can be quantized, giving the following form of the partition function
\begin{widetext}
	\begin{align}
		\mathcal{Z}&=\sum_{q=-\infty}^{\infty}\exp\bigg\{-\beta\|c\|^2\bigg[2KL_rL^2\pi^2\sum_{i=1}^{4}\Sigma_i(h)q^i-\frac{q^2}{2}\mu_B\bigg]\bigg\}\cr
		&\qquad\qquad\qquad\qquad\times\sum_{n=0}^{\infty}\delta(n)\exp\bigg\{-2\beta KL_r \|c\|^2\pi^2n\sum_{i=1}^{4}\tilde{\Sigma}_i(h)q^i\bigg\},
	\end{align}
\end{widetext}
where $n$ and the sum over it derive from the quantization of $e_0$ and $\delta(n)$ is the degeneracy on each state $n$. Once more, we consider the conditions $L_\phi=L_\theta$ and $p=q$.


\section{Baryonic tubes coupled to an external magnetic field} \label{SpaghettiSec}


In the previous sections of this paper, we defined the analytical solutions of baryonic layers immersed in an external, constant magnetic field using the \textit{lasagna ansatz}. In this section, we guess whether this is also possible when the \textit{spaghetti ansatz} is used. Unfortunately, we did not succeed in finding analytical solutions to the Skyrme equations. Indeed, as we are going to explain in the following sections, the terms introduced by the external field prevented us from decoupling the Skyrme equations. Any tentative simplification (for instance, by imposing new conditions of the external field of another ansatz on the Skyrme field parameters) led to incompatibilities of the Skyrme equations. This precludes us from giving a satisfactory analysis of the physics of baryonic tubes under the action of an external field, but we were able to outline some important properties of the main quantities, such as the role of the Skyrme term and the behavior of the energy density. 

As it is known from \cite{Skyrme1,Skyrme2,SkyrmeI,SkyrmeIII,SkyrmeIII+,tHooft1NExp,WittenBaryons1NExp,ANW,Canfora1:HedgehogAnsatz,Canfora2:NonlinearSuperposition,Canfora3:ChimPot,Canfora4:AdSWormholes,Canfora5:4dEinstein-nonlinear,Canfora6:U1gauged,Canfora7:Ordered(2018),Canfora8:Lasagne,Canfora9:AnalyticChristals,Canfora10:TraversableNUT-AdSWormholes,Canfora11:SuperCond,Canfora12:Pion,Canfora13:Gauged(May2021),Canfora14:YM,SergioFabrizio:2020zui,NostroI,NostroII}, the exponential parameterization is suitable to describe baryonic tubes. We recall here the \textit{spaghetti ansatz}, defined by
\begin{align}
	U(t,r,\theta,\phi)&= \exp (\chi (r) \tau_1),  \label{exponential}
\end{align}
where $\tau_1 = \vec n \cdot \vec T=n_1 T_1+n_2 T_2+n_3 T_3$. The ansatz is specified by
\begin{gather}\label{SA1}
	\vec n =(\sin \Theta \cos \Phi, \sin \Theta
	\sin \Phi, \cos \Theta) \\\label{SA2}
	\Theta =q\theta ,\quad \Phi =p\left( \frac{t}{L_\phi}-\phi \right) , \quad
	 p, q\in \mathbb{N}\ .
\end{gather}
The matrices $T_i$ define a basis of a three-dimensional subalgebra of $\mathfrak{g}=\mbox{Lie}(G)$. They can be normalized in such a way as to satisfy
\begin{align}
	[T_j,T_k]=\varepsilon_{jkm} T_m,\qquad\mathrm{Tr} (T_jT_k)=-2 I_{G,\rho} \delta_{jk},
\end{align}
where $I_{G,\rho}$ is called \textit{the Dynkin index} (see \cite{Dynkin:1957um,NostroII}). This ansatz satisfies the \textit{orthogonality conditions}
\begin{gather}\label{CondSpaghetti}
	\partial_\mu\Phi\partial^\mu\Phi=\partial_\mu\Phi\partial^\mu\Theta=\partial_\mu\Phi\partial^\mu\chi=\partial_\mu\Theta\partial^\mu\chi=0,
\end{gather}
which are necessary to reduce the system of Skyrme equations to one ODE. With a direct computation, one can show that the introduction of an external field along $T=T_3$ contributes through the shift
\begin{align}
	\partial_\mu\Phi\rightarrow\partial_\mu\Phi+A_\mu^{\mbox{ext}}.
\end{align}
Let us call $\mathcal{A}_\mu = A_\mu^{\mbox{ext}}+\partial_\mu\Phi$. It is worth noticing here that this shift \textit{breaks} part of the conditions \eqref{CondSpaghetti}. Indeed, $\mathcal{A}_\mu\mathcal{A}^\mu\neq 0$ (before the translation, $\partial_\mu\Phi$ was a null vector). This does not allow us to uncouple the equations of $\Theta$ and $\chi$ and, as a consequence, $\partial_\mu\chi\partial^\mu\Theta\neq 0$. For this reason, it is not possible to consider the two functions depending on different space-time variables; thus, let us define $\chi=\chi(r,\theta)$ and $\Theta=\Theta(r,\theta)$. One can try to simplify the equations by choosing $A_\mu^{\mbox{ext}}$ in different ways. For instance, the conditions 
\begin{gather}\label{CondSpaghettiA}
	\mathcal{A}_\mu\partial^\mu\Theta=\mathcal{A}_\mu\partial^\mu\chi=0.
\end{gather}
allow keeping part of the conditions \eqref{CondSpaghetti}. The explicit form of the external field is given by
\begin{align}\label{PotSpTheta}
	A_\mu^{\mbox{ext}}=(0,0,0,A_\phi(r,\theta))^T,
\end{align}
which defines a magnetic field along $\theta$ and $r$. Notice that, in this case, the magnetic field lies perpendicular to the tubes; thus, one can expect that it breaks the symmetry of the system.  A second possibility is to choose $A_\mu^{\mbox{ext}}$ in such a way to define a magnetic field along the tubes. For example,
\begin{align}\label{PotSpPhi}
	A_\mu^{\mbox{ext}}=(0,0,A_\theta(r),0)^T.
\end{align}
Notice that in both cases the external field has been chosen in such a way that 
\begin{align}\label{dA0}
     \partial_\mu\mathcal{A}^\mu=0.
\end{align}
The problem with this choice lies in the fact that all the conditions \eqref{CondSpaghetti} can be no longer satisfied, but it leads to an advantage. Indeed, as discussed better below, we want to conserve the baryonic tube shape of the soliton; in this scope, the energy density and the baryon density should be constant in $\phi$: we can always choose $\Phi=p(t/L_\phi-\phi)$ and one of the three Skyrme equations must become trivial; otherwise, it is not possible to find a solution to the system. 

In the following sections, we will not analyze both these situations in detail, but we will describe the general problem and the properties for $\lambda=0$ and $\lambda\neq 0$. Moreover, in Appendix \ref{AppFlavorOscillating} we propose an alternative form of the external field, in which its direction in the Lie algebra of the flavor group depends on the space-time coordinates. In this case, it becomes really simple to reduce the Skyrme equations to an analytically solvable ODE (in particular, the usual universal solution appears). We called it \textit{The flavor oscillating spaghetti}. In any case, a better characterization of the baryonic tube solution requires, probably, a better understanding of the spaghetti ansatz geometry.


\subsection{Baryonic tubes solutions in the NL$\sigma$M}


As already defined above, the Skyrme model with $\lambda=0$ defines the NL$\sigma$M. The general equations can be written as (see Appendix \ref{AppGeneralEq})
\begin{align}\label{INLS}
	&\partial_\mu\partial^\mu\chi-\sin\chi\left(\partial_\mu\Theta\partial^\mu\Theta+\sin^2\Theta\mathcal{A}_\mu\mathcal{A}^\mu\right)=0,\\ \label{IINLS}
	&(1-\cos\chi)\partial_\mu\partial^\mu\Theta+\sin\chi\partial_\mu\chi\partial^\mu\Theta\cr
	&\qquad\qquad\quad-
	(1-\cos\chi)\sin\Theta\cos\Theta\mathcal{A}_\mu\mathcal{A}^\mu=0,\\ \label{IIINLS}
	&\sin\chi\sin^2\Theta\partial_\mu\chi\mathcal{A}^\mu\cr
	&\qquad\qquad+2(1-\cos\chi)\sin\Theta\cos\Theta\partial_\mu\Theta\mathcal{A}^\mu=0,
\end{align}
where only the condition \eqref{dA0} on $\mathcal{A}_\mu$ has been considered (in particular, it is independent on the choice \eqref{PotSpTheta} and \eqref{PotSpPhi}. We can make the following observations.

\textit{First of all}, to get baryonic tubes, we can use the ansatz $\Phi=p\left( t/L_\phi-\phi \right)$. Indeed, this cancels the dependence from $\phi$ in the energy density. As already mentioned, the other functions cannot take the form of the ansatz  \eqref{SA1}, since the terms introduced by the external field \textit{couples} the equations, but we can impose the following orthogonality conditions
\begin{align}
    \partial_\mu\chi\partial^\mu\Phi=\partial_\mu\Theta\partial^\mu\Phi=0,
\end{align}
which means that $\chi$ and $\Theta$ are functions of only $r$ and $\theta$. In this case, the energy density is
	{\small 
		\begin{align}
			\rho_E=& K\|c\|^2\bigg\{\mathcal{A}_\mu\mathcal{A}^\mu(1-\cos\chi)\sin^2\Theta\cr 
            &\qquad+\frac{1}{2}\partial_\mu\chi\partial^\mu\chi+(1-\cos\chi)\left(\partial_\mu\Theta\partial^\mu\Theta\right)\bigg\}.
		\end{align}
	}

\textit{Secondly}, when the ansatz \eqref{PotSpTheta} is applied, the last equation \eqref{IIINLS} is automatically satisfied, due to the conditions \eqref{CondSpaghetti} and \eqref{dA0}. On the other hand, when \eqref{PotSpPhi} is applied, all the three equations are non-trivial, but one of them should be \textit{redundant}, since the equation depends only on $\chi$ and $\Theta$.

\textit{Thirdly}, when the external gauge field becomes big, the equation \eqref{INLS}, \eqref{IINLS} and \eqref{IIINLS} loose coercivity. This fact has also been encountered in the baryonic layers case. We are going to show that the Skyrme term saves the consistency of the Skyrme equations and also for the spaghetti ansatz.

It is worth mentioning here that the energy density diverges for a growing external field. This is predictable since the external fields contribute ti the energy of the systems. This behavior has also been observed for the baryonic layers. The main difference here consists in the fact that the solutions of the equations of motion necessarily depend on the external field (we need to remember that the baryonic layers allow for a \textit{universal} solution, independent from the external field). Thus, one may expect that the tube configuration's energy density (and so the total energy) diverges faster than the layers configuration's energy. If this is the case, the baryonic tubes become less stable than the baryonic layers and a configuration transition could appear. Those types of transitions were also considered in \cite{NostroI}.


\subsection{Baryonic tubes solutions in the Skyrme model}


The contribution of the Skyrme term to the general equations for the exponential parameterization is reported in Appendix \ref{AppGeneralEq}, where the condition \eqref{dA0} can be applied. It is straightforward to show that the first two observations considered in the previous section for the NL$\sigma$M remain valid. The main advantage of the introduction of the Skyrme term consists in the fact that the high limit of the external field can now be considered. With a direct computation, one can observe that the Skyrme equations take the following form 
\begin{widetext}
	{\small\begin{align}\label{LI}
			&\sin\chi\sin^2\Theta\bigg\{\mathcal{A}_\mu\mathcal{A}^\mu+\frac{\lambda}{4}
			\left[(\partial_\mu\chi\partial^\mu\chi)(\mathcal{A}_\nu\mathcal{A}^\nu)-(\partial_\mu\chi\mathcal{A}^\mu)^2\right]+\lambda(1-\cos\chi)\left[(\partial_\mu\Theta\partial^\mu\Theta)(\mathcal{A}_\nu\mathcal{A}^\nu)-(\partial_\mu\Theta\mathcal{A}^\mu)^2\right]\bigg\}\cr
			&\qquad\qquad-\frac{\lambda}{2}\partial_\mu\left[(1-\cos\chi)\sin^2\Theta(\mathcal{A}_\nu\mathcal{A}^\nu)\partial^\mu\chi\right]+\frac{\lambda}{2}\partial_\mu\left[(1-\cos\chi)\sin^2\Theta(\partial_\nu\chi\mathcal{A}^\nu)\mathcal{A}^\mu\right]
			=0,
	\end{align}}
	
	{\small\begin{align}\label{LII}
			&(1-\cos\chi)\sin\Theta\cos\Theta\bigg\{\mathcal{A}_\mu\mathcal{A}^\mu+\frac{\lambda}{4}
			\left[(\partial_\mu\chi\partial^\mu\chi)(\mathcal{A}_\nu\mathcal{A}^\nu)-(\partial_\mu\chi\mathcal{A}^\mu)^2\right]+\frac{\lambda}{2}(1-\cos\chi)\left[(\partial_\mu\Theta\partial^\mu\Theta)(\mathcal{A}_\nu\mathcal{A}^\nu)-(\partial_\mu\Theta\mathcal{A}^\mu)^2\right]\bigg\}\cr
			&\qquad\qquad-\frac{\lambda}{2}\partial_\mu\left[(1-\cos\chi)^2\sin^2\Theta(\mathcal{A}_\nu\mathcal{A}^\nu)\partial^\mu\Theta\right]+\frac{\lambda}{2}\partial_\mu\left[(1-\cos\chi)\sin^2\Theta(\partial_\nu\Theta\mathcal{A}^\nu)\mathcal{A}^\mu\right]
			=0,
	\end{align}}
	
	{\small\begin{align}\label{LIII}
			&\partial_\mu\bigg\{\left[(1-\cos\chi)\sin^2\Theta\mathcal{A}^\mu\right]
			+\frac{\lambda}{4}\left[(1-\cos\chi)\sin^2\Theta(\partial_\nu\chi\partial^\nu\chi)\mathcal{A}^\mu\right]-\frac{\lambda}{4}\left[(1-\cos\chi)\sin^2\Theta(\partial_\nu\chi\mathcal{A}^\nu)\partial^\mu\chi\right]\cr
			&\qquad\qquad+\frac{\lambda}{2}\left[(1-\cos\chi)^2\sin^2\Theta(\partial_\nu\Theta\partial^\nu\Theta)\mathcal{A}^\mu\right]-\frac{\lambda}{2}\left[(1-\cos\chi)^2\sin^2\Theta(\partial_\nu\Theta\mathcal{A}^\nu)\partial^\mu\Theta\right]\bigg\}
			=0.
	\end{align}}
\end{widetext}
Also in this case, we tried different ways to find solutions to these equations, but we did not succeed.

As for the previous case, the energy may be considered growing faster than the solution with the lasagna ansatz. Thus, a transition between the two configurations could happen. The point of transition could depend on the parameter $\lambda$. In this case, the energy density takes the general form

\begin{widetext}
	{\small 
		\begin{align}
			\rho_E=& K\|c\|^2\bigg\{\mathcal{A}_\mu\mathcal{A}^\mu(1-\cos\chi)\sin^2\Theta\left[1+\frac{\lambda}{2}(1-\cos\chi)\partial_\mu\Theta\partial^\mu\Theta+\frac{\lambda}{4}\partial_\mu\chi\partial^\mu\chi\right]+\frac{1}{2}\partial_\mu\chi\partial^\mu\chi\cr
			&
			\qquad\qquad\qquad\qquad\qquad\qquad\qquad\qquad+(1-\cos\chi)\left[\partial_\mu\Theta\partial^\mu\Theta+\frac{\lambda}{4}\left(\partial_\mu\chi\partial^\mu\chi\partial_\nu\Theta\partial^\nu\Theta-(\partial_\mu\chi\partial^\mu\Theta)^2\right)\right]\bigg\}.
		\end{align}
	}
\end{widetext}

\section{Discussion and perspectives}
We have considered a comparison between the Skyrme model and the NL$\sigma$M (which is represented by the Skyrme model with $\lambda=0$) in the presence of an external electromagnetic field. We used both the exponential and the Euler parametrization for the ansatz. Interestingly, in both cases, the equations of motion for the pure
NL$\sigma$M become unsolvable when the external field is very strong (tends to infinity). Instead, for $\lambda\neq0$, the equations of motion are always solvable. Specifically, the Euler parameterization ansatz reduces the Skyrme equations to an ODE and keeps the layer's shape. This determines a universal solution, that is independent of the external field, therefore existing in all cases when the external field is zero, finite, or infinite. For baryonic tubes, the situation is quite different since the equations of motion are difficult to be solved analytically. However, it seems that introducing an external field hampers the existence of solutions that keep the baryonic tube structure, leading to incompatible sets of equations. One may wonder why these topological solutions disappear in the limit when $\lambda=0$. The answer is not new in nonlinear equations and can be read from equation \eqref{FirstLinearSol}. We see that the solution is proportional to the inverse square root of the coupling constant, so compensating the coupling in the interactions and giving finite energy contribution terms independent on $\lambda$. \\
In particular, we concentrated on the thermodynamics of the universal solution, when also a possibly complex Baryonic chemical potential is included. Very interestingly, a quite rich Stokes phenomenon structure appears, manifesting different phases of the Baryonic structure. Interestingly, such a manifestation of resurgence appears in such a concrete model. In particular, our calculations come back to the Stokes phenomenology for Pearcey integrals in all possible cases. To this end, we have approximated infinite sums with integrals. It will be interesting in future work to study the Stokes phenomenon directly for the series. However, we have reasons to think of a priory that we should not expect any qualitative changes w.r.t. the integral case. Nevertheless, such computations could become interesting for more precise quantitative predictions. To this end, however, we need first to further improve our model by including some corrections we neglected here.\\
One of the most important issues is that we neglected back reactions in the presence of the external field. For example, consider \eqref{vediamo}. This expression represents the energy density inside the box. Assuming the topological configuration is not destroyed, only gauge transformations compatible with the boundary conditions are allowed for the internal field, while arbitrary gauge transformations are allowed for the external gauge field. This means that an additional contribution must compensate for the effect of the external field through polarization, through the distribution of charges along the boundary, and then generate a compensating internal field. We did not include such polarization effects here, to avoid further technical complications to already involved mathematical computations. We mean to study them separately in future work. \\
Another issue deserving attention regards the solutions in \ref{LasagnaSol}, for $\lambda/L_r^2<16$.
In this case, the first derivative in $r$ is discontinuous and the second derivative gives delta distributions at the discontinuities. Stokes' theorem thus requires the insertion of brane sources located at the discontinuities. This kind of solution therefore deserves a more detailed study by including brane sources' contributions. Here, we limited our analysis to the case reported in Appendix \ref{App:Piecewise}, but further details going beyond the aim of the paper will be the subject of future work.\\
Finally, similar considerations as above can be done for the analysis we presented in Sec. \ref{sec:suscept}. Here the problem is that the back-reaction may probably give an important contribution to $F$ and $\chi$ also for small values of $h$, since it is expected to be of the order of the external field. Therefore, it is not really negligible. This issue deserves further investigation, presently under consideration. Furthermore, we preferred to leave the analysis of these physical quantities in terms of the temperature for future work, with the aim of including also the contribution back-reaction. Indeed, as already discussed above, the internal reaction of the skyrmion is important in order to determine the value of the external field that breaks our solutions, but also to study the possible phase transitions and the behavior of the critical temperature in terms of the external field.

\section*{Acknowledgements}
We thank Prabal Adhikari for the relevant discussions. \\
F. C. has been funded by FONDECYT Grant 1240048.

\newpage

\appendix



\section{The flavor-oscillating spaghetti} \label{AppFlavorOscillating}


In this section, we want to show how the baryonic tubes can easily survive in an external field when the direction in the Lie algebra of the flavor group of the external field depends on the position in space-time. Indeed, as discussed in the paper, the choice of $T=T_3$ causes a translation of $\partial_\mu\Phi\rightarrow\partial_\mu\Phi-A_\mu^{\mbox{ext}}$. Since $\partial_\mu\Phi$ is a light-like vector, which has been chosen to cancel some terms in the ungauged case, this translation causes the presence of those terms. A solution could be the choice of a direction that does not lead to this type of translation. An example is provided $T=\tau_3$, where $\tau_3$ derives from the notations introduced in \cite{NostroI,NostroII}. Namely,
\begin{align}
	\tau_1&=\sin\Theta(\cos\Phi T_1+\sin\Phi T_2)+\cos\Theta T_3,\\
	\tau_2&=\partial_\Theta\tau_1=\cos\Theta(\cos\Phi T_1+\sin\Phi T_2)-\sin\Theta T_3,\\
	\tau_3&=\frac{\partial_\Phi\tau_1}{\sin\Theta}=-\sin\Phi T_1+\cos\Phi T_2.
\end{align}
Remember that $\tau_1$ has also been used in the definition of $U$ in \eqref{exponential}. This way, the contribution of the external field to the left current is accounted for
\begin{align}
	\hat{\mathcal{L}}_\mu=U^{-1}D_\mu U=\mathcal{L}_\mu-A_\mu^{\mbox{ext}} U^{-1}\left[\tau_3,U\right].
\end{align}
With a direct computation, one finds that
\begin{align}
	U^{-1}\left[\tau_3,U\right]=\sin\chi\tau_2-(1-\cos\chi)\tau_3.
\end{align}
On the other hand,
\begin{align}
	&\mathcal{L}_\mu=\partial_\mu\chi\tau_1 + \partial_\mu\Phi\sin\Theta\left[\sin {\chi }\tau_{3}+(1-\cos {\chi }
	)\tau_{2}\right]\cr 
	&\qquad\qquad\qquad+ \partial_\mu\Theta\left[\sin {\chi }\tau_{2}-(1-\cos {\chi })\tau_{3}\right].
\end{align}
Thus,  the external field causes a translation of $\partial_\mu\Theta$. Let us call $\mathcal{A}_\mu=A_\mu^{\mbox{ext}}-\partial_\mu\Theta$. Using the conditions 
\begin{align}
	\partial_\mu\Phi\partial^\mu\Phi=0\andd A_\mu^{\mbox{ext}}\partial^\mu\Phi=A_\mu^{\mbox{ext}}\partial^\mu\chi=0,
\end{align} 
the equations are simplified to
\begin{align}
	&\partial_{\mu}\partial^{\mu}\chi\left[1+\frac{\lambda}{2}\mathcal{A}_\nu\mathcal{A}^\nu\left(1-\cos\chi\right)\right]\cr
	&\qquad\qquad-\mathcal{A}_\mu\mathcal{A}^\mu\sin\chi\left(1-\frac{\lambda}{4}\partial_\nu\chi\partial^\nu\chi\right)=0,\\
	&\partial_{\mu}\mathcal{A}^\mu=0\andd\partial_{\mu}\partial^\mu\Phi=0.
\end{align}
The above conditions and the last two equations are automatically satisfied with the usual choices
\begin{gather}
	A_\mu^{\mbox{ext}}=(0,0,A_\theta^{\mbox{ext}}(r),0),\\
	\Phi=\Phi(u)\andd\Theta=q\theta.
\end{gather}
This way, one can choose $\chi=\chi(r)$ and reduce the equations to a second-order ODE
\begin{align}
	&\chi''\left[1+\frac{\lambda}{2}\mathcal{A}_\nu\mathcal{A}^\nu\left(1-\cos\chi\right)\right]\cr
	&\qquad\qquad-\mathcal{A}_\mu\mathcal{A}^\mu\sin\chi\left(1-\frac{\lambda}{4L_r^2}\chi'^2(r)\right)=0,
\end{align}
which admits a universal solution, independent of the value of the external field.

The main problem linked to this type of description is the fact that the direction of the external field in the space of the Lie algebra of the flavor group depends on the space-variable $\phi$. Usually, the coefficients of the $T_i$ is the definition of $\tau_1$ (which in the ansatz \eqref{exponential} have been called $n_i$, as defined in \eqref{SA1}) are recognized with the three charged pions, $\pi^\pm$ and $\pi^0$ \cite{Skyrme1,Skyrme2,SkyrmeI,SkyrmeIII,SkyrmeIII+,ANW}. It is easy to see that when $T=T_3$, the neutral pion corresponds to $n_3$. In this case, the flavor of the neutral pion \textit{oscillates}, since the direction of the external field \textit{oscillates}. This is a strange fact and, for this reason, we reported the treatment of this model here in the Appendix, excluding it from the "official" computations.


\section{On the equations of motion.} \label{App:EqMotion}
For the baryonic tubes, a simple calculation gives
\begin{align}
 U^{-1}\partial_\mu U=&\partial_\mu \chi \tau_1+\sin \chi \partial_\mu n^j T_j\cr &+(1-\cos\chi) \partial_\mu n^j n^k \epsilon_{jkl}T_l.
\end{align}
Similarly,
\begin{align}
 U^{-1}T_3 U-T_3=&(1-\cos \chi)(n^3 n^jT_j-T_3)\cr &+\sin\chi \epsilon_{3jk} n^jT_k.
\end{align}
After some manipulations we then get
\begin{align}
 \hat {\mathcal L}=& \partial_\mu \chi \tau_1+\sin \chi \Big(\partial_\mu n^j-\epsilon_{3kj} n^k A^{\rm ext}_\mu\Big) T_j \cr
 &+(1-\cos\chi) \Big( \partial_\mu n^j -\epsilon_{3kj} n^k A^{\rm ext}_\mu \Big) n^h \epsilon_{jhl}T_l.
\end{align}
On the other hand, we have also 
\begin{align}
 \partial_\mu n^j=\partial_\mu \Theta \partial_\Theta n^j+\partial_\mu \Phi \epsilon_{3kj}n^k,
\end{align}
so that 
\begin{widetext}
\begin{align}
  \hat {\mathcal L}=& \partial_\mu \chi \tau_1+\sin \chi \Big(\partial_\mu \Theta \partial_\Theta n^j+\epsilon_{3kj} n^k(\partial_\mu\Phi- A^{\rm ext}_\mu)\Big) T_j 
  +(1-\cos\chi) \Big( \partial_\mu \Theta \partial_\Theta n^j+\epsilon_{3kj} n^k(\partial_\mu\Phi- A^{\rm ext}_\mu) \Big) n^h \epsilon_{jhl}T_l.
\end{align}
\end{widetext}
Therefore, we can write $ \hat {\mathcal L}=\mathcal L [\partial_\mu\Phi-A^{\rm ext}_\mu]$. Hence, for 
\begin{align}
 V^\mu\equiv \hat{\mathcal{L}}^\mu-\frac{\lambda}{4}\left[\hat{G}^{\mu\nu},\hat{\mathcal{L}_\nu}\right],
\end{align}
we can write the equations of motion as
\begin{align}
 \partial_\mu V^\mu +[\hat{\mathcal{L}}_\mu,V^\mu]=A^{\rm ext}_\mu[T_3,V^\mu].
\end{align}
The l.h.s. is thus obtained easily from the equations of motion without external field, by shifting $\partial_\mu\Phi$, while in the r.h.s., $V^\mu$ is still a function of $\partial_\mu\Phi-A^{\rm ext}_\mu$ but the extra term $A^{\rm ext}_\mu$ seems to
break this scheme.\\
However, we can prove that this is not the case in the following way. The point is that in $\partial_\mu V^\mu$ we have to do the shift before taking the derivative, but since $V^\mu$ is a functional of $\Phi$, through $n^j$, after deriving we get new $\partial_\mu \Phi$
terms. We show that, because of gauge invariance of the action, these new terms pair with the r.h.s. of the above equations giving once again the usual shift. To this aim, let us first explore a bit the gauge transformations. These are given by
\begin{align}
 U&\longmapsto e^{\alpha T_3} U e^{-\alpha T_3}, \\
 A^{\rm ext}_\mu &\longmapsto A^{\rm ext}_\mu+\partial_\mu\alpha.
\end{align}
Applying this to $\hat{\mathcal L}$ gives $e^{\alpha T_3} \hat{\mathcal L}e^{-\alpha T_3}$, so the lagrangian is gauge invariant.
An elementary calculation shows that the gauge transformation of $U$ simply gives the translation $\Phi\longmapsto \Phi+\alpha$. This explains why $A_\mu$ appears in the combination $\partial_\mu\Phi-A^{\rm ext}_\mu$ in $\hat{\mathcal L}$: it is the only
possible gauge invariant combination. Now, let us write the functional dependence of $V^\mu$:
\begin{align}
 V^\mu=V^\mu[\chi,\Theta,\Phi,\partial_\mu\chi,\partial_\mu\Theta,\partial_\mu\Phi-A^{\rm ext}_\mu].
\end{align}
Since the combination $\partial_\mu\Phi-A^{\rm ext}_\mu$ is gauge invariant, and given the effect of gauge transformations, we can write
\begin{align}
 V^\mu=&e^{\Phi T_3}V^\mu[\chi,\Theta,0,\partial_\mu\chi,\partial_\mu\Theta,\partial_\mu\Phi-A^{\rm ext}_\mu] e^{-\Phi T_3}\cr
 =&e^{\Phi T_3}V^\mu_0 e^{-\Phi T_3}.
\end{align}
Therefore,
\begin{align}
 \partial_\mu V^\mu=& e^{\Phi T_3}\partial_\mu V^\mu_0 e^{-\Phi T_3}+e^{\Phi T_3}\partial_\mu \Phi [T_3,V^\mu_0] e^{-\Phi T_3},
\end{align}
and so
\begin{align}
 \partial_\mu V^\mu&-A^{\rm ext}_\mu[T_3, V^\mu]= e^{\Phi T_3}\partial_\mu V^\mu_0 e^{-\Phi T_3}\cr &+e^{\Phi T_3}(\partial_\mu\Phi -A^{\rm ext}_\mu) [T_3,V^\mu_0] e^{-\Phi T_3},
\end{align}
as we wanted to prove. This shows that the equations gauged with $A^{\rm ext}_\mu$ are obtained from the ungauged ones by shifting $\partial_\mu$ by $-A^{\rm ext}_\mu$ everywhere.  

\

The same calculations work for the baryonic layer solutions. In this case a gauge transformation shifts $\Phi\to \Phi+\beta$, $\Theta\to \Theta-\beta$, so only the combination $\alpha$ is affected by the gauge transformation, with
$\alpha \to \alpha-\beta$. Therefore, in this case, it necessarily must appear everywhere in the gauge invariant combination $\partial_\mu \alpha+A^{\rm ext}_\mu$.


\section{General form of the Skyrme equations} \label{AppGeneralEq}


In this section, we report the general form of the Skyrme equations (i.e., with $\lambda\neq 0$) for both the Euler and exponential parameterization.


\subsection{The Euler parameterization}


In the case of Euler parameterization, we remember that the Skyrme field is written as
\begin{align}\label{EulerMapApp}
	U=e^{\Phi\kappa}e^{\chi h}e^{\Theta\kappa}.
\end{align}
Since we are considering the general equations, without taking any ansatz on the function representing the Euler angles, $\chi$, $\Theta$, and $\Phi$ depend in a generalized way on the space-time variables. The only ansatz is taken on the matrices $\kappa$ and $h$, which are the same as the ones defined in Section \ref{LasagnaSection}. As deeply analyzed in \cite{NostroII}, the map \eqref{EulerMapApp} describes a closed cycle when the boundary conditions on $\Phi$, $\Theta$ and $\chi$ are chosen in such a way that 
\begin{gather}\label{Ranges}
	0\leq\Phi\leq \eta\sigma {2\pi},\quad
	0\leq\Theta\leq \eta {2\pi},\\
	0\leq\chi\leq n{\pi},
\end{gather}
where $\sigma=\eta=1$ for odd-dimensional representations of the elements of the Lie algebra $\mathfrak{g}$ and $\sigma=\frac{1}{2}$ and $\eta=2$ for even-dimensional representations (which means that $\sigma=\frac{1}{\eta}$).

With these choices, the Skyrme equations take the form
\begin{widetext}
	{\small 
		\begin{align}
			&\partial_{\mu}\partial^{\mu}\chi \left\{1+\lambda\left[%
			\mathcal{A}_\mu\mathcal{A}^\mu\sin^2\left(\frac{\chi}{2}%
			\right)+\partial_{\mu}\xi\partial^{\mu}\xi\cos^2\left(\frac{\chi}{2}\right)%
			\right]\right\} -\sin(\chi)\left(1-\frac{\lambda}{4}\partial_{\mu}\chi%
			\partial^{\mu}\chi\right)\left(\mathcal{A}_{\nu}\mathcal{A}^{\nu}-%
			\partial_{\nu}\xi\partial^{\nu}\xi\right) \cr & -\lambda\sin(\chi)\cos(%
			\chi)\left[\mathcal{A}_{\mu}\mathcal{A}^{\mu}\partial_{\nu}\xi%
			\partial^{\nu}\xi-\left(\mathcal{A}_{\mu}\partial^{\mu}\xi\right)^2\right]
			-\lambda\left\{\sin^2\left(\frac{\chi}{2}\right)\partial_{\mu}\mathcal{A}^{%
				\mu}\mathcal{A}_{\nu}\partial^{\nu}\chi+\cos^2\left(\frac{\chi}{2}%
			\right)\partial_{\mu}\partial^{\mu}\xi\partial_{\nu}\xi\partial^{\nu}\chi%
			\right.\cr &+\sin^2\left(\frac{\chi}{2}\right)\left[\mathcal{A}_{\mu}%
			\partial^{\mu}\left(\mathcal{A}_{\nu}\partial^{\nu}\chi\right)-\partial_{%
				\mu}\chi\partial^{\mu}\left(\mathcal{A}_{\nu}\mathcal{A}^{\nu}\right)%
			\right] \left.+\cos^2\left(\frac{\chi}{2}\right)\left[\partial_{\mu}\xi%
			\partial^{\mu}\left(\partial_{\nu}\xi\partial^{\nu}\chi\right)-\partial_{%
				\mu}\chi\partial^{\mu}\left(\partial_{\nu}\xi\partial^{\nu}\xi\right)\right]%
			\right\}\cr & -\frac{\lambda}{4}\sin(\chi)\left[\left(\mathcal{A}_{\mu}%
			\partial^{\mu}\chi\right)^2-\left(\partial_{\mu}\xi\partial^{\mu}\chi%
			\right)^2\right]=0,  \label{FirstSkyrme}
		\end{align}
	}	
	{\small 
		\begin{align}
			&4\sin\left(\frac{\chi}{2}\right)\Bigl\{ \sin\left(\frac{\chi}{2}\right)%
			\Bigl\{ \partial_{\mu}\mathcal{A}^{\mu}\left[1+\frac{\lambda}{4}%
			\partial_{\nu}\chi\partial^{\nu}\chi\right] -\frac{\lambda}{4}\left[%
			\partial_{\mu}\partial^{\mu}\chi\partial_{\nu}\chi\mathcal{A}^{\nu}+%
			\partial_{\mu}\chi\partial^{\mu}\left(\partial_{\nu}\chi\mathcal{A}^{\nu}%
			\right)-\mathcal{A}{\mu}\partial^{\mu}\left(\partial_{\nu}\chi\partial^{%
				\nu}\chi\right)\right] \Bigr\} \cr & +\cos\left(\frac{\chi}{2}\right)%
			\Bigl\{ \frac{\lambda}{2}\sin(\chi)\partial_{\mu}\mathcal{A}^{\mu}%
			\mathcal{A}_{\nu}\partial^{\nu}\xi -\frac{\lambda}{2}\sin(\chi)\left[%
			\partial_{\mu}\partial^{\mu}\xi\mathcal{A}_{\nu}\mathcal{A}^{\nu}+%
			\partial_{\mu}
			\xi\partial^{\mu}\left(\mathcal{A}_{\nu}\mathcal{A}^{\nu}\right)-%
			\mathcal{A}_{\mu}\partial^{\mu}\left(\mathcal{A}_{\nu}\partial^{\nu}%
			\xi\right)\right]\cr &+\lambda\cos(\chi)\left[\partial_{\mu}\chi%
			\mathcal{A}^{\mu}\mathcal{A}_{\nu}\partial^{\nu}\xi-\partial_{\mu}\chi%
			\partial^{\mu}\xi\mathcal{A}_{\nu}\mathcal{A}^{\nu}\right]%
			+b\partial_{\mu}\chi\partial^{\mu}\xi \Bigr\} \Bigr\}\cr &-4\cos\left(\frac{\chi%
			}{2}\right)\Bigl\{ \cos\left(\frac{\chi}{2}\right)\Bigl\{ %
			\partial_{\mu}\partial^{\mu}\xi\left[1+\frac{\lambda}{4}%
			\partial_{\nu}\chi\partial^{\nu}\chi\right] -\frac{\lambda}{4}%
			\left[\partial_{\mu}\partial^{\mu}\chi\partial_{\nu}\chi\partial^{\nu}\xi+%
			\partial_{\mu}\chi\partial^{\mu}\left(\partial_{\nu}\chi\partial^{\nu}\xi%
			\right)-\partial_{\mu}\xi\partial^{\mu}\left(\partial_{\nu}\chi\partial^{%
				\nu}\chi\right)\right] \Bigr\}\cr &+\sin\left(\frac{\chi}{2}\right)\Bigl\{ 
			\frac{\lambda}{2}\sin(\chi)\partial_{\mu}\partial^{\mu}\xi\mathcal{A}_{\nu}%
			\partial^{\nu}\xi-\frac{\lambda}{2}\sin(\chi)\left[%
			\partial_{\mu}\mathcal{A}^{\mu}\partial_{\nu}\xi\partial^{\nu}\xi+%
			\mathcal{A}_{\mu}\partial^{\mu}\left(\partial_{\nu}\xi\partial^{\nu}\xi%
			\right)-\partial_{\mu}\xi\partial^{\mu}\left(\mathcal{A}_{\nu}\partial^{%
				\nu}\xi\right)\right]\cr & +\lambda\cos(\chi)\left[\partial_{\mu}\chi%
			\partial^{\mu}\xi\mathcal{A}_{\nu}\partial^{\nu}\xi-\partial_{\mu}\chi%
			\mathcal{A}^{\mu}\partial_{\nu}\xi\partial^{\nu}\xi\right]%
			-\partial_{\mu}\chi\mathcal{A}^{\mu} \Bigr\} \Bigr\}=0,
			\label{SecondSkyrme}
		\end{align}
	}	
	{\small 	
		\begin{align}
			&4\sin\left(\frac{\chi}{2}\right)\left\{ \cos\left(\frac{\chi}{2}\right)%
			\Bigl\{ \partial_{\mu}\mathcal{A}^{\mu}\left[1+\frac{\lambda}{4}%
			\partial_{\nu}\chi\partial^{\nu}\chi\right] -\frac{\lambda}{4}\left[%
			\partial_{\mu}\partial^{\mu}\chi\partial_{\nu}\chi\mathcal{A}^{\nu}+%
			\partial_{\mu}\chi\partial^{\mu}\left(\partial_{\nu}\chi\mathcal{A}^{\nu}%
			\right)-\mathcal{A}{\mu}\partial^{\mu}\left(\partial_{\nu}\chi\partial^{%
				\nu}\chi\right)\right] \Bigr\}\right.\cr & -\sin\left(\frac{\chi}{2}\right)%
			\Bigl\{  \frac{\lambda}{2}\sin(b\chi)\partial_{\mu}\mathcal{A}^{\mu}%
			\mathcal{A}_{\nu}\partial^{\nu}\xi -\frac{\lambda}{2}\sin(\chi)\left[%
			\partial_{\mu}\partial^{\mu}\xi\mathcal{A}_{\nu}\mathcal{A}^{\nu}+%
			\partial_{\mu}
			\xi\partial^{\mu}\left(\mathcal{A}_{\nu}\mathcal{A}^{\nu}\right)-%
			\mathcal{A}_{\mu}\partial^{\mu}\left(\mathcal{A}_{\nu}\partial^{\nu}%
			\xi\right)\right]\cr &+\lambda\cos(\chi)\left[\partial_{\mu}\chi%
			\mathcal{A}^{\mu}\mathcal{A}_{\nu}\partial^{\nu}\xi-\partial_{\mu}\chi%
			\partial^{\mu}\xi\mathcal{A}_{\nu}\mathcal{A}^{\nu}\right]%
			+\partial_{\mu}\chi\partial^{\mu}\xi \Bigr\} \Bigr\}\cr & +4\cos\left(\frac{%
				\chi}{2}\right)\Bigl\{ \sin\left(\frac{\chi}{2}\right)\Bigl\{ %
			\partial_{\mu}\partial^{\mu}\xi\left[1+\frac{\lambda}{4}%
			\partial_{\nu}\chi\partial^{\nu}\chi\right] -\frac{\lambda}{4}%
			\left[\partial_{\mu}\partial^{\mu}\chi\partial_{\nu}\chi\partial^{\nu}\xi+%
			\partial_{\mu}\chi\partial^{\mu}\left(\partial_{\nu}\chi\partial^{\nu}\xi%
			\right)-\partial_{\mu}\xi\partial^{\mu}\left(\partial_{\nu}\chi\partial^{%
				\nu}\chi\right)\right]\Bigr\} \cr &-\cos\left(\frac{\chi}{2}\right)%
			\Bigl\{ \frac{\lambda}{2}\sin(\chi)\partial_{\mu}\partial^{\mu}\xi\mathcal{A}_{\nu}%
			\partial^{\nu}\xi-\frac{\lambda}{2}\sin(\chi)\left[%
			\partial_{\mu}\mathcal{A}^{\mu}\partial_{\nu}\xi\partial^{\nu}\xi+%
			\mathcal{A}_{\mu}\partial^{\mu}\left(\partial_{\nu}\xi\partial^{\nu}\xi%
			\right)-\partial_{\mu}\xi\partial^{\mu}\left(\mathcal{A}_{\nu}\partial^{%
				\nu}\xi\right)\right]\cr & +\lambda\cos(\chi)\left[\partial_{\mu}\chi%
			\partial^{\mu}\xi\mathcal{A}_{\nu}\partial^{\nu}\xi-\partial_{\mu}\chi%
			\mathcal{A}^{\mu}\partial_{\nu}\xi\partial^{\nu}\xi\right]%
			-\partial_{\mu}\chi\mathcal{A}^{\mu} \Bigr\} \Bigr\}=0,
			\label{ThirdSkyrme}
		\end{align}
	} 
\end{widetext}
with $\alpha=\frac{1}{2}(\Theta-\Phi)$, $\xi=\frac{1}{2}(\Theta+\Phi)$.


\subsection{The exponential parameterization}


The general equations for the exponential parameterization are obtained considering the following form for the Skyrme field
\begin{align}
	U(t,r,\theta,\phi)&= \exp (\chi \tau_1),
\end{align}
with, as defined in Section \ref{SpaghettiSec},
\begin{gather}
	\tau_1 = \vec n \cdot \vec T=n_1 T_1+n_2 T_2+n_3 T_3\\
	\vec n =(\sin \Theta \cos \Phi, \sin \Theta
	\sin \Phi, \cos \Theta).
\end{gather}
As for the baryonic layers, here, $\chi$, $\Theta$, and $\Phi$ are general functions of the space-time variables. Here, the boundaries are defined by
\begin{gather}\label{Ranges1}
	0\leq\Phi\leq {2p\pi},\quad
	0\leq\Theta\leq {q\pi},\\
	0\leq\chi\leq n{\pi},
\end{gather}
with $n$, $p$ and $q$ integers. The Skyrme equations are
\begin{widetext}
	{\small\begin{align}\label{LLI}
			&\partial_\mu\partial^\mu\chi-\sin\chi\left(\partial_\mu\Theta\partial^\mu\Theta+\sin^2\Theta\mathcal{A}_\mu\mathcal{A}^\mu\right)\cr
			&-\frac{\lambda}{4}\bigg\{
			\sin\chi\left[(\partial_\mu\chi\partial^\mu\chi)(\partial_\nu\Theta\partial^\nu\Theta)-(\partial_\mu\chi\partial^\mu\Theta)^2\right]
			+\sin\chi\sin^2\Theta\left[(\partial_\mu\chi\partial^\mu\chi)(\mathcal{A}_\nu\mathcal{A}^\nu)-(\partial_\mu\chi\mathcal{A}^\mu)^2\right]\cr
			&+4\sin\chi(1-\cos\chi)\sin^2\Theta\left[(\partial_\mu\Theta\partial^\mu\Theta)(\mathcal{A}_\nu\mathcal{A}^\nu)-(\partial_\mu\Theta\mathcal{A}^\mu)^2\right]\cr
			&-2\partial_\mu\left[(1-\cos\chi)(\partial_\nu\Theta\partial^\nu\Theta)\partial^\mu\chi\right]+2\partial_\mu\left[(1-\cos\chi)(\partial_\nu\chi\partial^\nu\Theta)\partial^\mu\Theta\right]\cr
			&-2\partial_\mu\left[(1-\cos\chi)\sin^2\Theta(\mathcal{A}_\nu\mathcal{A}^\nu)\partial^\mu\chi\right]+2\partial_\mu\left[(1-\cos\chi)\sin^2\Theta(\partial_\nu\chi\mathcal{A}^\nu)\mathcal{A}^\mu\right]
			\bigg\}=0,
	\end{align}}
	
	{\small\begin{align}\label{LLII}
			&(1-\cos\chi)\partial_\mu\partial^\mu\Theta+\sin\chi\partial_\mu\chi\partial^\mu\Theta-(1-\cos\chi)\sin\Theta\cos\Theta\mathcal{A}_\mu\mathcal{A}^\mu\cr
			&-\frac{\lambda}{4}\bigg\{
			(1-\cos\chi)\sin\Theta\cos\Theta\left[(\partial_\mu\chi\partial^\mu\chi)(\mathcal{A}_\nu\mathcal{A}^\nu)-(\partial_\mu\chi\mathcal{A}^\mu)^2\right]\cr
			&+2(1-\cos\chi)^2\sin\Theta\cos\Theta\left[(\partial_\mu\Theta\partial^\mu\Theta)(\mathcal{A}_\nu\mathcal{A}^\nu)-(\partial_\mu\Theta\mathcal{A}^\mu)^2\right]\cr
			&-\partial_\mu\left[(1-\cos\chi)(\partial_\nu\chi\partial^\nu\chi)\partial^\mu\Theta\right]+\partial_\mu\left[(1-\cos\chi)(\partial_\nu\chi\partial^\nu\Theta)\partial^\mu\chi\right]\cr
			&-2\partial_\mu\left[(1-\cos\chi)^2\sin^2\Theta(\mathcal{A}_\nu\mathcal{A}^\nu)\partial^\mu\Theta\right]+2\partial_\mu\left[(1-\cos\chi)\sin^2\Theta(\partial_\nu\Theta\mathcal{A}^\nu)\mathcal{A}^\mu\right]
			\bigg\}=0,
	\end{align}}
	
	{\small\begin{align}\label{LLIII}
			&\partial_\mu\bigg\{\left[(1-\cos\chi)\sin^2\Theta\mathcal{A}^\mu\right]
			+\frac{\lambda}{4}\left[(1-\cos\chi)\sin^2\Theta(\partial_\nu\chi\partial^\nu\chi)\mathcal{A}^\mu\right]-\frac{\lambda}{4}\left[(1-\cos\chi)\sin^2\Theta(\partial_\nu\chi\mathcal{A}^\nu)\partial^\mu\chi\right]\cr
			&+\frac{\lambda}{2}\left[(1-\cos\chi)^2\sin^2\Theta(\partial_\nu\Theta\partial^\nu\Theta)\mathcal{A}^\mu\right]-\frac{\lambda}{2}\left[(1-\cos\chi)^2\sin^2\Theta(\partial_\nu\Theta\mathcal{A}^\nu)\partial^\mu\Theta\right]\bigg\}
			=0.
	\end{align}}
\end{widetext}

\section{The caustic curve}\label{app:hmu}
Since the parameters $\eta_1, \eta_3$ depend linearly on the magnetic field $h$, we find convenient to introduce the constants $\tilde \eta_i$, $i=1,3$, so that
\begin{align}
    \eta_1=-h\tilde \eta_1, \qquad\ \eta_3=-h\tilde \eta_3.
\end{align}
If we further introduce $y\equiv y(\mu,h)$ by
\begin{align}
    y=\eta_4 \tilde \eta_2 -\frac 3{32} \tilde \eta_3^2 h^2,\label{ipsilon}
\end{align}
because of \eqref{tildeeta}, $y$ is linear in $\mu_B$, and the cubic \eqref{CubicDiscriminant} takes the form
\begin{align}
    y^3&-\frac {27}8 h^2 \tilde\eta_3 \left(\tilde \eta_1 \eta_4^2 -\frac {\tilde \eta_3^3 h^2}{2^7}\right)y\cr
    &+\frac {27}8 \left(h^2 \tilde \eta_1^2\eta_4^4+\frac {5}{32} \tilde \eta_1\tilde \eta_3^3\eta_4^2 -\frac {\tilde \eta_3^6}{2^{11}}  \right)=0. \label{takestheform}
\end{align}
The discriminant of this cubic is
\begin{align}
    \Delta_y=\frac {3^6}{2^{20}} h^4 (16 \tilde \eta_1 \eta_4^2-\tilde \eta_3^3 h^2)^3.
\end{align}
It is positive for small $h$, negative for large $h$, changes sign at (recalling $\tilde \eta_j>0$ for $j=1,3$)
\begin{align}
    \bar h=4 \sqrt {\frac {\tilde \eta_1}{\tilde \eta_3}} \frac {\eta_4}{\tilde \eta_3},
\end{align}
and vanishes at $h=0$. For $h<\bar h$ there is only one branch, while a second branch appears at $h\geq \bar h$. Since there are no further zeros of $\Delta_y$,
the two branches cannot intersect. To understand the shape of the caustic in the $(h,y)$ plane, it is convenient to determine explicitly an expression for the solutions at $h\sim 0$, $h\to \infty$ and $h\sim \bar h$.\\
For $h=0$ the unique solution is $y=0$ (triple degenerate). For small $h$, one then immediately finds that the unique solution is
\begin{align}
    y_{0} = -\frac 32 \tilde \eta_1^{\frac 23} \eta_4^{\frac 43} h^{\frac 23}+O(h^2). \label{ipsilon0}
\end{align}
For the other cases, it is convenient to use Cardano's formula
\begin{align}
    y_i=\omega^{i-1} \left(-\frac q2 +\sqrt {\Delta_y}\right)^{\frac 13} + \bar \omega^{i-1} \left(-\frac q2 -\sqrt {\Delta_y}\right)^{\frac 13},
\end{align}
$i=1,2,3$, where $\omega=\frac 12 (-1+i\sqrt 3)$, and we have written \eqref{takestheform} in the form $y^3+py+q=0$. For very large $h$ we have
\begin{align}
    q&\approx -\frac {27}{2^{14}} \tilde \eta_3^6 h^6, \\
    \Delta_y&\approx -\frac {3^6}{2^{20}} \tilde \eta_3^9 h^{10}.
\end{align}
Therefore, $\sqrt{|\Delta_x|}$ is smaller than $q$ and we can write
\begin{align}
    y_i\approx \omega^{i-1} \left(-\frac q2 \right)^{\frac 13} \left(1 -\frac {2i\sqrt {|\Delta_y|}}{3q}\right) + c.c., \label{approx}
\end{align}
where $c.c.$ means complex conjugate. More explicitly:
\begin{align}
    y_{\infty,1}&= \frac 3{16} \tilde \eta_3^2 h^2+O(1), \\
    y_{\infty,2}&= -\frac 3{32} \tilde \eta_3^2 h^2 \left(1-\frac {32}{\tilde \eta_3^{\frac 32}\sqrt 3 h} \right)+O(1),\\
    y_{\infty,3}&= -\frac 3{32} \tilde \eta_3^2 h^2 \left(1+\frac {32}{\tilde \eta_3^{\frac 32}\sqrt 3 h} \right)+O(1).
\end{align}
To understand to which branch this solution belongs, it is sufficient to analyze the solutions in $h>\bar h$, very near to $\bar h$.
Since now $\Delta_y$ is very small, the expressions \eqref{approx} are still valid but with different values of $q$ and $\Delta_y$:
\begin{align}
    \frac q2&\approx \left(\frac {9\tilde \eta_1 \eta_4^2}{2\tilde \eta_3}\right)^3, \\
    \Delta_y&\approx -\frac {3^6}{2^{17}} \bar h^3 \tilde \eta_3^3 (h-\bar h)^3.
\end{align}
Therefore, we get
\begin{align}
    y_{\bar h,1}&= -\frac{9\tilde \eta_1 \eta_4^2}{\tilde \eta_3}+O((h-\bar h)^2), \\
    y_{\bar h,2}&= \frac{9\tilde \eta_1 \eta_4^2}{2\tilde \eta_3} \left(1-\frac {\sqrt 2 \tilde \eta_3^{\frac 92}\bar h^{\frac 32}}{\sqrt 3 6^6 \tilde \eta_1^3\eta_4^6}(h-\bar h)^{\frac 32} \right)+O((h-\bar h)^2),\\
    y_{\bar h,3}&= \frac{9\tilde \eta_1 \eta_4^2}{2\tilde \eta_3} \left(1+\frac {\sqrt 2 \tilde \eta_3^{\frac 92}\bar h^{\frac 32}}{\sqrt 3 6^6 \tilde \eta_1^3\eta_4^6}(h-\bar h)^{\frac 32} \right)+O((h-\bar h)^2).
\end{align}
We see that $y_{\bar h,2}$ and $y_{\bar h,3}$ coincide at $h=\bar h$ so belong to the new branch. $y_{\bar h,1}$ necessarily belongs to the branch of $y_0$. Since
the two branches cannot intersect and $y_{\bar h,1}$ is below $y_{\bar h,2}$ and $y_{\bar h,3}$, we infer that $y_{\infty,3}$ belongs to the branch starting from $h=0$,
while $y_{\infty,2}$ belongs to the lower part of the second branch and $y_{\infty,1}$ belongs to the upper part of the second branch. We considered the branch 
starting at $\bar h$ as a unique branch, being connected, but we remark that its upper and lower part form a cusp in $h=\bar h$, since there
\begin{align}
   \left. \frac {dy_{\bar h,2}}{dh} \right|_{h=\bar h}=\left. \frac {dy_{\bar h,3}}{dh} \right|_{h=\bar h}=0.
\end{align}
From this analysis, we can easily reconstruct the picture in the $(h,\mu_B)$-plane simply by using \eqref{ipsilon} and \eqref{tildeeta}. It then follows that the 
simple branch results to be above while the double branch is below. Notice also that from \eqref{ipsilon0} and \eqref{tildeeta} we get for the simple branch that it 
starts from
\begin{align*}
    \mu_B=4KL_rL^2\pi^2\ \Sigma_2(h,m) \left(\frac{3\pi}{L_\theta^2}+\frac{6\pi}{L_\phi^2}\right)
\end{align*}
with an infinite positive slope. Finally, we can notice that the caustic is necessarily confined in the region where $b<0$. By writing
$\eta_2=\eta_2^{(0)}+\eta_2^{(1)}h^2$, where $\eta_2^{(i)}$ are positive constants, we see that such condition corresponds to
\begin{align}
    \frac {\|c\|^2}2 \mu_B \geq \eta_2^{(0)}+\left(\eta_2^{(1)}-\frac 38 \frac {\tilde \eta_3^2}{\eta_4}\right) h^2.
\end{align}
From the expressions relating the $\eta$s to the $\Sigma(h,m)$s, we see that the parenthesis is positive when
\begin{align}
    \Sigma_2^{(1)} \Sigma_4-\frac 38 \tilde \Sigma_3^2>0,
\end{align}
where analogously to the $\eta$s, we have defined $\Sigma_3=:\tilde \Sigma_3 h$ and $\Sigma_2:=\Sigma_2^{(0)}+\Sigma_2^{(1)}h^2$. From the expressions
\eqref{eta1}--\eqref{eta5}, we see that the positiveness is never satisfied.\\
Finally, it is interesting to notice that the lowest component of the double branch is definitely decreasing. Indeed, from the expression for $y_{\infty,1}$ we get
\begin{align}
    \frac {\|c\|^2}2 \mu_{B,\infty,1}=\eta_2^{(0)}+\left(\eta_2^{(1)}-\frac {9}{32} \frac {\tilde \eta_3^2}{\eta_4}\right) h^2.
\end{align}
As above, for $m\geq 1$ the expression in parenthesis results to be always negative, therefore, for large $h$ such a lower component becomes negative.\\
Thus, we see that the qualitative shape of the caustic in the $(h,\mu_B)$ plane is universal, it does not depend on the specific values of the parameters. 
A representative picture is shown in Figure \ref{Fig:muBVSB}.

\section{The Pearcey integral} \label{App:pearcey}
Here we recall the main properties of the Pearcey integral \eqref{PIntegral}, expressed in the variable $x$ and $y$, introduced in \cite{Paris91}, related to the standard $X$ and $Y$ 
variables by $x=Xe^{-i\frac \pi4}$, $y=Ye^{i\frac \pi8}$. For any fixed value of $x$ and $y$, we can expand the exponentials $e^{-xq^2}$ and $e^{iyq}$ in power
series to get
\begin{align}
    \mathcal{P}(x,y)=\frac 12 e^{i\frac \pi8} \sum_{n=0}^\infty \sum_{m=0}^\infty \frac {(-1)^m}{m!(2n)!} \Gamma\left(\frac {m+n}2+\frac 14\right) x^my^{2n},
\end{align}
which converges in any compact polydisc. Therefore, the interesting cases arise when $x$, or $y$, or both become very large. Even if a complete treatment is still lacking, these cases are well-studied in the literature, and we report here the main results.


\textbf{Large $x$ expansion.} 
An exhaustive analysis of the asymptotic behavior of the Pearcey integral for large $x$ at any fixed value for $y$ can be found in \cite{Paris91}.
One has to distinguish two cases. For $|\arg (x)|\leq\frac{\pi}{2}$ one finds the complete asymptotic expansion
\begin{align}
    \mathcal{P}(x,y)\sim& \sqrt{\frac \pi{x}} e^{i\frac \pi8} S_1(x,y), \\
    S_1(x,y)=& e^{-\frac {y^2}{4x}} \sum_{m=0}^\infty \frac {(-1)^m a_m(y^2/4x)}{m! x^{2m}}, \\
    a_m(z)=& 2^{-4m} H_{4m}(\sqrt{z}),
\end{align}
where $H_n$ is the $n$-th Hermite polynomial.\\
For $|\arg (-x)|<\frac{\pi}{2}$ one finds
\begin{align}
    \mathcal{P}(x,y)\sim& \sqrt{\frac \pi{x}} e^{i\frac \pi8} \left[\sigma(x) S_1(x,y)+i\sqrt{2} e^{\frac {x^2}4} S_2(x,y)\right], \label{E5} \\
    \sigma(x)=& -{\rm sign}\arg(-x), \\
    S_2(x,y)=& \sum_{m=0}^\infty \frac {(y^2/8x)^{2m}}{m!}\Big[P(2m,\xi)\cos\xi\cr
    & \phantom{\sum_{m=0}^\infty \frac {(y^2/8x)^{2m}}{m!}} -Q(2m,\xi)\sin\xi\Big], \\
    P(2m,\xi)=& \sum_{k=0}^m \frac {(-1)^k (2m+2k)!(2\xi)^{-2k}}{(2k)! (2m-2k)!},\\
    Q(2m,\xi)=& \sum_{k=0}^{m-1} \frac {(-1)^k (2m+2k+1)!(2\xi)^{-2k-1}}{(2k+1)! (2m-2k-1)!},\\
    \xi=&y\left( -\frac x2 \right)^{\frac 12}.
\end{align}
The main ingredient here is the exponential $e^{\frac {x^2}4}$ in \eqref{E5}. From it, we see that when $|\arg (-x)|<\frac{\pi}{4}$ the Pearcey integral grows 
exponentially far away from the origin, while after crossing the half-lines $\arg (x)=\pm \frac 34 \pi$ it oscillates with a dominating algebraic behavior. 
Such half-lines are
Stokes lines in the complex $x$ plane, independently on $y$, as $y$ stays bounded. A constant discontinuity appears at $\arg (x)=\pi$, and for $\arg(z)=\pm \frac \pi2$
we have a transition from a double function contribution ($S_1+S_2$) to a single function contribution ($S_1$): these three half-lines are anti-Stokes 
lines\footnote{It is worth remarking here that our convention on Stokes lines is the opposite of that in \cite{Paris91}}.

\textbf{Large $y$ expansion.} 
The case of bounded $x$ and large $|y|$ is carefully developed in \cite{Pagola}, \footnote{We agree with their conventions on Stokes lines}. Since the Pearcey 
integral is even in $y$, it is sufficient to consider the region $\arg (y)\leq \frac \pi2$. In this case, one gets
\begin{widetext}
\begin{align}
    \mathcal P(x,y)=\begin{cases}
    P_1(x,y)+P_2(x,y) +O(e^{-1.38077|y|^{\frac 43}}) & {\mbox{ if }}\qquad|\arg (y)|\leq \frac \pi8, \\
    P_1(x,y) & {\mbox{ if }}\qquad -\frac \pi2 \leq |\arg (y)|< -\frac \pi8, \\
    P_2(x,y) & {\mbox{ if }}\qquad \frac \pi8 < |\arg (y)|\leq \frac \pi2,
    \end{cases}
\end{align}
where
\begin{align}
    P_k(x,y)\sim& \frac {\sqrt{\pi/3}}{2^{\frac 56} y^{\frac 13}} \exp \left[ 3 \left( \frac y4\right)^{\frac 43} e^{-(-1)^{k} 2i\frac \pi3} 
    -x\left( \frac y4\right)^{\frac 23} e^{-(-1)^{k} i\frac \pi3} +\frac {x^2}6\right] 
    \left[ \sum_{n=0}^m e^{(-1)^k(2n+1)i\frac \pi6} \frac {A_n(x)}{y^{\frac 23n}}+O(\frac 1{y^{2m+3}})  \right],
\end{align}
and
\begin{align}    
    A_n(x)=&\sum_{r=\lfloor \frac{n+1}2 \rfloor}^n \sum_{l=0}^{2r-n}(-1)^{n+l} a_{n,r,l}(x) c_{2r+n-l}(x), \\
    a_{n,r,l}(x) =&\frac {(-1)^r x^l 2^{\frac 43 (2r-n-l)}}{k! (2r-n-l)! (n-r)!}, \\
    c_{2r+n-l}(x)=&\frac {x^n}{3^n 2^{\frac n3}} \sum_{k=0}^{\lfloor \frac{n}2 \rfloor} \left( \frac 3{2x^2} \right)^k \frac {n!}{k!(n-2k)!}.
\end{align}
\end{widetext}
From this, one can infer that there is a Stokes line along the real half-line $\arg (y)=0$, and two anti-Stokes lines at $\arg(y)=\pm \frac 38\pi$.

\begin{figure}[!htbp]
\begin{center}
\resizebox{7.5cm}{!}{
\begin{tikzpicture}[>=latex]  
\draw [thick] (-5,0)--(5,0);
\draw [thick] (0,-2)--(0,5);
\draw [ultra thick] (-5,0.1) .. controls (-2,0.1) and (-0.1,2) .. (-0.1,5);
\draw [ultra thick] (5,0.1) .. controls (2,0.1) and (0.1,2) .. (0.1,5);
\draw [ultra thick, red] (0,0) -- (1.73205,0);
\draw [ultra thick,->,red] (0.5,0) -- (0.7,0);
\draw [ultra thick,->,red] (1.4,0) -- (1.2,0);
\draw [ultra thick, red] (-2,0) -- (-1.73205,0);
\draw [thick,->,red] (-1.85,0) -- (-2,0);
\draw [ultra thick, dashed] (1,-2)--(1,5);
\draw [ultra thick,->] (-1.43,1.43)--(-1.23,1.63);
\node at (-1.7,1.7){$\pmb {\Gamma_2}$};
\draw [ultra thick,-<] (1.43,1.43)--(1.23,1.63);
\node at (1.7,1.7){$\pmb {\Gamma_1}$};
\filldraw (0,0) circle (1.5pt);
\node at (-0.2,-0.2){$\pmb 0$};
\filldraw (1,0) circle (1.5pt);
\node at (0.75,-0.3){$\pmb {\frac 1{\sqrt 6}}$};
\filldraw (1.73205,0) circle (1.5pt);
\node at (1.73205,-0.3){$\pmb {\frac 1{\sqrt 2}}$};
\filldraw (-1.73205,0) circle (1.5pt);
\node at (-1.53205,0.3){$\pmb {-\frac 1{\sqrt 2}}$};
\filldraw (-2,0) circle (1.5pt);
\node at (-2.3,-0.4){$\pmb {-\sqrt {\frac 23}}$};
\end{tikzpicture}
}
\end{center}
\caption{The integration path is deformed from the real axis to the union of the two curves $\Gamma_1$ and $\Gamma_2$. The red lines indicate the displacing of the real solutions $z_1<0\leq z_2\leq z_3$ when $\phi$ varies from $0$ to $\pi/6$.}\label{IntCont}
\end{figure}
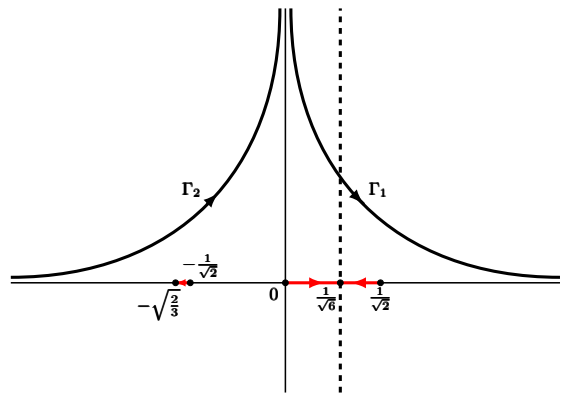

\textbf{Large parameters near the caustic.} 
This case is considered in \cite{Kaminski}. The analysis, despite being valid also for certain complex value, are not as complete as in the previous case and allows us to do 
just a partial analysis, which is however interesting since it concerns the case when $x$ and $y$ are real. The key fact is that in such a situation the number of saddle
points contributing to the asymptotic expansion depends on the sign of the discriminant $\Delta$: when positive, only one saddle point contributes, while when negative,
three saddle points contribute. The delicate question is what happens at the caustic. Since the discriminant can change sign only for negative $b$, it follows that it is
sufficient to consider the case of large negative $x$. To use the results in \cite{Kaminski}, it is worth mentioning that there is a further different
convention for the Pearcey integral. If we call $P(X,Y)$ the standard definition and $P_K(X,Y)$ the one in \cite{Kaminski}, then we have
\begin{align}
    \mathcal P(Xe^{-i\frac \pi4},Ye^{i\frac \pi8})=P(X,Y)=\frac 1{\sqrt 2} P_K(X,\frac {Y}{\sqrt 2}).
\end{align}
The analysis in \cite{Kaminski} (se also \cite{Stamnes}) is done for real $X,Y$, while we are interested in real $x$ and purely imaginary $y$, which correspond to complex $X$ and $Y$. 
In particular, we are exactly at the boundary of the region of the validity of their analysis, see the comment below formula (2.2) in \cite{Kaminski}.
Since we are not sure to stress their results to the boundary, it is convenient for us to repeat the calculation in our case.\\
As in \cite{Kaminski}, the interesting situation is near the points where the caustic changes sign, and, when $x$ is real, also in our case, this happens when $x$ is
large and negative. Thus, we change $x\to -x$ and consider $x$ positive. The caustic is along $y^2=\frac {8}{27}x^3$, so we are interested in the large $x$ behaviour
of
\begin{align}
    Q(x,\mu)=\mathcal P(-x,\mu x^{\frac 32}),
\end{align}
where we set
\begin{align}
    \mu=\sqrt {\frac {8}{27}}-\alpha,
\end{align}
and $\alpha$ is a small real parameter. When $\alpha$ is positive there are three real critical points, while when $\alpha$ is negative there are one real and two complex conjugate critical points. It is convenient to do the calculations for positive $\alpha$ and then consider analytic continuation.
After a simple change of coordinates, we can write
\begin{align}
    Q(x,\mu)=e^{i\frac \pi8}\sqrt x \int_{\mathbb R} e^{-x^2(z^4-z^2+\mu z)}dz.
\end{align}
The critical points of the quartic polynomial in $z$ are
\begin{align}
    z_1=&-\sqrt {\frac 23} \sin \left(\frac \pi3+\phi\right),\\
    z_2=&-\sqrt {\frac 23} \sin \left(\phi\right),\\
    z_3=&\sqrt {\frac 23} \sin \left(\frac \pi3-\phi\right),
\end{align}
where $\phi$ is such that $|\phi|\leq \frac \pi6$ and is defined by
\begin{align}
    \phi=\frac 13 \arcsin\left(\mu \sqrt {\frac {27}8}\right). 
 \end{align}
Notice that when $\alpha\to 0$ then $\phi\to \frac \pi6$ and the two positive roots coincide at $z_2=z_3=1/\sqrt 6$. Taking into account this fact, it is convenient to deform the integration contour to $\Gamma=\Gamma_1\cup\Gamma_2$, where $\Gamma_j$ are depicted in Fig \ref{IntCont}.

\begin{figure}[!htbp]
\begin{center}
\resizebox{8cm}{!}{
\begin{tikzpicture}[>=latex]  
\filldraw [gray!8!white] (-3,0)--(7.5,10.5*0.414213562)--(7.5,7.5)--(-7.5,7.5)--(-7.5,4.5*0.414213562)--cycle;
\filldraw [gray!8!white] (-3,0)--(7.5,-10.5*0.414213562)--(7.5,-7.5)--(-7.5,-7.5)--(-7.5,-4.5*0.414213562)--cycle;
\filldraw [gray!20!white] (-3,0)--(7.5,10.5*0.414213562)--(7.5,-10.5*0.414213562)--cycle;
\filldraw [gray!20!white] (-3,0)--(-7.5,4.5*0.414213562)--(-7.5,-4.5*0.414213562)--cycle;
\filldraw [gray!20!white] (-3,0)--(-3+7.5*0.414213562,7.5)--(-3-7.5*0.414213562,7.5)--cycle;
\filldraw [gray!20!white] (-3,0)--(-3+7.5*0.414213562,-7.5)--(-3-7.5*0.414213562,-7.5)--cycle;
\draw [thick] (-8,0)--(8,0);
\draw [thick] (-3,-8)--(-3,8);
\draw [ultra thick,gray!50!black] (-3-7.5*0.414213562,-7.5)--(-3+7.5*0.414213562,7.5);
\draw [ultra thick,gray!50!black] (-3-7.5*0.414213562,7.5)--(-3+7.5*0.414213562,-7.5);
\draw [ultra thick,gray!50!black] (-7.5,-4.5*0.414213562)--(7.5,10.5*0.414213562);
\draw [ultra thick,gray!50!black] (-7.5,4.5*0.414213562)--(7.5,-10.5*0.414213562);
\draw [ultra thick,dashed] (-3+6.5*0.414213562,-7.5)--(-3+6.5*0.414213562,7.5);
\node at (-2.5,7.8) {${\rm Im}(t)$};
\node at (7.8,-0.3) {${\rm Re}(t)$};
\draw [ultra thick, red] (-7.5,0)--(7.5,0);
\draw [ultra thick, blue] (-3,0) .. controls (-3,4) and (-5.5,5.5) .. (-7.5,7.5);
\draw [ultra thick, blue] (-3,0) .. controls (-3,-4) and (-5.5,-5.5) .. (-7.5,-7.5);
\draw [ultra thick, blue] (1.5,0) .. controls (1.5,4) and (4,5.5) .. (6,7.5);
\draw [ultra thick, blue] (1.5,0) .. controls (1.5,-4) and (4,-5.5) .. (6,-7.5);
\draw [ultra thick, blue] (1,0) .. controls (1,2) and (-2.9+6.5*0.414213562,3.5) .. (-2.9+6.5*0.414213562,7.5);
\draw [ultra thick, blue] (1,0) .. controls (1,-2) and (-2.9+6.5*0.414213562,-3.5) .. (-2.9+6.5*0.414213562,-7.5);
\filldraw (-3,0) circle (2pt);
\filldraw (-3+6.5*0.414213562,0) circle (2pt);
\filldraw (1,0) circle (2pt);
\filldraw (1.5,0) circle (2pt);
\node at (-3.3+6.5*0.414213562,-0.2){$\pmb {-z_1}$};
\node at (0.8,-0.2){$\pmb {z_2}$};
\node at (1.7,-0.2){$\pmb {z_3}$};
\end{tikzpicture}
}
\end{center}
\caption{The red line represent the steepest descent line passing through $z_1$ (i.e. $t=0$). The blue lines represent real curves for the exponent. The dark region represents valleys at large 
$t$ while the light regions are the hills.}\label{curves}
\end{figure}
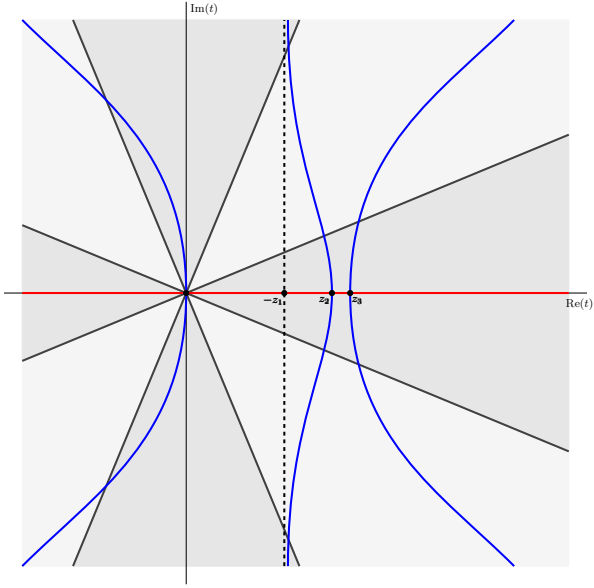

From the picture, we see that in this way, we can always deform $\Gamma_1$ so that it passes through $z_2$ and $z_3$, while $\Gamma_1$ passes through $z_1$. It is convenient to separate things in this manner to have better control of what happens when the two positive critical points merge.\\
We first analyze the integral along $\Gamma_1$. Following \cite{Kaminski}, it is convenient to consider a sequence of changes of variables. First, we introduce the shift $z=t+\frac 1{\sqrt 6}$ so that the exponent becomes
\begin{align}
    z^4-z^2+\mu z=:g(t,\alpha)+\frac 19-\frac {\alpha}{\sqrt 6},
\end{align}
with
\begin{align}
   g(t,\alpha)=t^4+\frac 4{\sqrt 6} t^3-\alpha t. 
\end{align}
Then, we introduce a second change of variable $u=u(t)$, defined by
\begin{align}
   g(t,\alpha)=\frac {u^3}3-\zeta u+\eta. \label{cubicu}
\end{align}
Here, we want the change of variables to be smooth at the points $t_j=z_j-\frac 1{\sqrt 6}$, $j=2,3$. Taking the derivative of the above expression w.r.t. $u$, we get
\begin{align}
    \partial_t (t,\alpha)\frac {dt}{du}=u^2-\zeta.
\end{align}
The only possibility for $\frac {dt}{du}$ to be finite and non-zero at $t_2$ and $t_3$
is that, correspondingly, $u=\pm \zeta^{\frac 12}$. We choose the plus sign to correspond to $t_3$. Therefore, we get
\begin{align}
    \zeta^{\frac 32}=& \frac 34 [g(t_2,\alpha)-g(t_3,\alpha)],\\
    \eta=& \frac 12 [g(t_2,\alpha)+g(t_3,\alpha)].
\end{align}
By solving \eqref{cubicu}, one then gets
\begin{align}
    u(t,\alpha)=& 2\zeta^{\frac 12} \sin \psi, \\
    \psi=\frac 13 \arcsin \frac {3(\eta-g(z,\alpha))}{2\zeta^{\frac 32}}.
\end{align}
From this one can show that the change of variable is analytic not only around the critical points but along the whole path $\Gamma_1-\frac 1{\sqrt 6}$. In the new coordinate $u$ the contribution of the $\Gamma_1$ curve to $Q$ is therefore
\begin{align}
    Q_1(x,\alpha)=e^{i\frac \pi8}\sqrt x P_1(x,\alpha), 
\end{align}
where for large values of $x$ we have
\begin{widetext}
    \begin{align}
        P_1(x,\alpha)\sim e^{-x^2\left( \frac 19-\frac \alpha{\sqrt 6}+\eta\right)} \sum_{n=0}^\infty x^{-2n} \left[p_n(\alpha) \int_C e^{-x^2 \left( \frac {u^3}3-\zeta u\right)}du + q_n(\alpha) \int_C ue^{-x^2 \left( \frac {u^3}3-\zeta u\right)}du\right],
    \end{align}
\end{widetext}
where $p_n$ and $q_n$ as exactly as defined in (3.7) of \cite{Kaminski}, and $C$ is the path starting at $e^{i\frac 23 \pi}\infty$ and ending at $+\infty$. The integrals along $C$ are easily expressed in terms of Airy functions so that we get
\begin{widetext}
    \begin{align}
        P_1(x,\alpha)\sim& -e^{-x^2\left( \frac 19-\frac \alpha{\sqrt 6}+\eta\right)} \frac {\pi}{x^{\frac 23}}\left( Bi (\zeta x^{\frac 43})+i Ai (\zeta x^{\frac 43}) \right) \sum_{n=0}^\infty x^{-2n} p_n(\alpha) \cr
        & -e^{-x^2\left( \frac 19-\frac \alpha{\sqrt 6}+\eta\right)} \frac {\pi}{x^{\frac 43}}\left( Bi' (\zeta x^{\frac 43})+i Ai' (\zeta x^{\frac 43}) \right) \sum_{n=0}^\infty x^{-2n} q_n(\alpha),
    \end{align}
\end{widetext}
where the prime indicates derivative w.r.t. the argument. In particular, we find
\begin{align}
    p_0(\alpha)=&\frac 1{\sqrt 2 3^{\frac 34}}+O(\alpha), \\
    q_0(\alpha)=&\frac 1{4\sqrt 2 3^{\frac 14}}+O(\alpha).
\end{align}
Now, let us consider the contribution of the path $\Gamma_2$. In this case, it is convenient to introduce the coordinate $t$ such that $z=t+z_1$ so that
\begin{align*}
    z^4-z^2+\mu z=z_1^4-z_1^2+\mu z_1 +\frac 12 t^2 (12z_1^2-2)+4t^3z_1+t^4.
\end{align*}
In looking for the steepest descent paths, after putting $t=x+iy$, we determine the curves where the polynomial in $t$ has zero imaginary part. This gives
\begin{align}
    0=-4y^3(x+z_1)+y(4x^3+12x^2z_1+x(12z_1^2-2)). 
\end{align}
This is easily solved, keeping into account that $z_1^2\leq \frac 12$. The curves are depicted in Fig. \ref{curves}.


We see that it is not possible to deform $\Gamma_2$ to the red curve since it would need to cross hills. However, we can deform it to the red curve until the point $-z_1$ and then to the 
upward dashed half-right line. In computing the saddle point approximation we then see that for large $x$ only the red line component contributes to the integral, while the remaining part gives
exponentially smaller results. The contribution of the $\Gamma_2$ curve to $Q$ is therefore
\begin{align}
    Q_2(x,\alpha)=e^{i\frac \pi8}\sqrt x P_2(x,\alpha), 
\end{align}
where
\begin{align}
    P_2(x,\alpha)\sim& e^{-x^2h(z_1)} \sqrt {\frac {\pi}{6z_1^2-1}}\Big[1+ \cr
    &\phantom{e^{-x^2h(z_1)}} +\frac 1{x^2}\left(15z_1^2-\frac 34\right)+\ldots\Big], \\
    h(z)=& z^4-z^2+\mu z.
\end{align}    
Putting it all together, we are now able to analyze the asymptotic behavior of the Pearcey integral around the caustic. To this aim, it is convenient to recall the following asymptotic expressions for the Airy functions (see \cite{abramowitz+stegun}):
\begin{align}
    Ai(s)\sim& \frac 1{2\sqrt\pi s^{\frac 14}} e^{-\xi}\sum_{k=0}^\infty (-1)^k \frac {c_k}{\xi^k}, \quad |\arg(z)|<\pi, \\
    Ai'(s)\sim& -\frac {s^{\frac 14}}{2\sqrt\pi} e^{-\xi}\sum_{k=0}^\infty (-1)^k \frac {d_k}{\xi^k}, \quad |\arg(z)|<\pi, \\
    Bi(s)\sim& \frac 1{\sqrt\pi s^{\frac 14}} e^{\xi}\sum_{k=0}^\infty \frac {c_k}{\xi^k}, \quad |\arg(z)|<\frac \pi3, \\
    Bi'(s)\sim& -\frac {s^{\frac 14}}{\sqrt\pi} e^{\xi}\sum_{k=0}^\infty \frac {d_k}{\xi^k}, \quad |\arg(z)|<\frac \pi3,     
\end{align}
where
\begin{align}
    \xi=\frac 23 s^{\frac 32},
\end{align}
and
\begin{align}
    c_0=&d_0=1, \qquad d_k=-\frac {6k+1}{6k-1}c_k, \\
    c_k=& \frac {\Gamma(3k+1/2)}{(54)^kk!\Gamma(k+1/2)}.
\end{align}
We will also need:
\begin{widetext}
    \begin{align}
        Bi(ze^{\pm \frac \pi3 i})\sim& \sqrt {\frac 2\pi}\frac {e^{\mp \frac \pi6 i}}{z^{\frac 14}} \Big[ \sin \Big( \xi+\frac \pi4 \mp \frac i2 \log2\Big) 
        \sum_{k=0}^\infty (-1)^k \frac {c_{2k}}{\xi^{2k}}-\cos \Big( \xi+\frac \pi4 \mp \frac i2 \log2\Big) 
        \sum_{k=0}^\infty (-1)^k \frac {c_{2k+1}}{\xi^{2k+1}}\Big] ,\\
        Bi'(ze^{\pm \frac \pi3 i})\sim& \sqrt {\frac 2\pi}e^{\mp \frac \pi6 i}z^{\frac 14} \Big[ \cos \Big( \xi+\frac \pi4 \mp \frac i2 \log2\Big) 
        \sum_{k=0}^\infty (-1)^k \frac {d_{2k}}{\xi^{2k}}+\sin \Big( \xi+\frac \pi4 \mp \frac i2 \log2\Big) 
        \sum_{k=0}^\infty (-1)^k \frac {d_{2k+1}}{\xi^{2k+1}}\Big] ,
    \end{align}
\end{widetext}
which are valid when $|\arg(z)|<\frac 23 \pi$. Let us write the final result in the form
\begin{widetext}
    \begin{align}
        Q(x,\alpha)\sim &-e^{i\frac \pi8}\sqrt x e^{-x^2\left( \frac 19-\frac \alpha{\sqrt 6}+\eta\right)}\Bigg[\frac{\pi}{\sqrt 6 3^{\frac 14}} \frac 1{x^{\frac 23}}
        (Bi(\zeta x^{\frac 43})+iAi(\zeta x^{\frac 43}))+\frac{3^{\frac 14}\pi}{4\sqrt 6} \frac 1{x^{\frac 43}}
        (Bi'(\zeta x^{\frac 43})+iAi'(\zeta x^{\frac 43}))\Bigg]\cr
        &+e^{i\frac \pi8}e^{-x^2h(z_1)} \sqrt{\frac {\pi}{6z_1^2-1}}\frac 1{\sqrt x}+\ldots.
    \end{align}
\end{widetext}
Since $h(z_1)<0$ for any $\alpha$, we see that the term in the second line ($Q_2(x,\alpha)$) is always dominating. The first line, instead, contributes in different ways according to the sign of $\alpha$. For $\alpha>0$, $\zeta$ is real, and using the above formulas for the Airy functions we get
\begin{align}
    Q_1(x,\alpha)\sim -\frac {e^{i\frac \pi8}}{\sqrt x}\sqrt {\frac{\pi}{6}} \frac {1-\sqrt{3\zeta}}{(3\zeta)^{\frac 14}}e^{-x^2h(z_3)}.
\end{align}
Here $h(z_3)$ is positive, so this term decays exponentially like a Gaussian.\\
For $\alpha<0$ instead oscillating modes appear. In this case, $\zeta=|\zeta|e^{i\frac \pi3}$ and we get
\begin{widetext}
   \begin{align}
   Q_1(x,\alpha)\sim&-\frac {e^{i\frac \pi8}}{\sqrt x}  e^{-x^2\left( \frac 19-\frac \alpha{\sqrt 6}+\eta\right)}
   \left[ \frac {\sqrt \pi e^{i\frac \pi4}}{3^{\frac 34}|\zeta|^{\frac 14}} \sin\left(\frac 23 |\zeta|^{\frac 32} x^2+\frac \pi4-\frac i2 \log 2\right) 
   -\frac {\sqrt \pi e^{-i\frac \pi{12}}|\zeta|^{\frac 14}}{ 3^{\frac 14}4} \cos\left(\frac 23 |\zeta|^{\frac 32} x^2+\frac \pi4-\frac i2 \log 2\right) \right]\cr
   & -\frac {e^{i\frac \pi8}}{\sqrt x}  e^{-x^2\left( \frac 19-\frac \alpha{\sqrt 6}+\eta\right)}
   i\frac {\sqrt \pi e^{-i\frac \pi{12}}}{2\sqrt 6 3^{\frac 14}} \left(1-\frac {\sqrt 6 |\zeta|^{\frac 12}e^{i\frac \pi6}}{4}\right) e^{-i\frac 23 |\zeta|^{\frac 32}x^2}.
   \end{align} 
\end{widetext}
Finally, for $\alpha=0$ we have $\zeta=0$ and $\eta=6^{-\frac 12}3^{-\frac 14}$, therefore we find
\begin{widetext}
    \begin{align}
    Q_1(x,\alpha)\sim&-e^{-i\frac \pi8}e^{-x^2\left( \frac 19+\frac 1{\sqrt 6 3^{\frac 14}}\right)} \frac {\pi}{\sqrt6 3^{\frac 14}}
    \left( \frac 1{3^{\frac 13}\Gamma(\frac 13)}+i\frac 1{3^{\frac 23}\Gamma(\frac 23)} \right)\frac 1{x^{\frac 16}}.
\end{align}
\end{widetext}


\section{The piecewise-linear solutions}\label{App:Piecewise}


In this section, we want to give an insight into the case when the condition ${\lambda}/{16 L_r^2}=1$ is not satisfied. 

The requirement to satisfy the imposed boundary conditions is
\begin{align}
	\frac{\lambda}{16L_r^2}<1.
\end{align}
With this condition, the universal solution takes the form
\begin{align}\label{SolPezzi}
	\chi(r)=\left\{\begin{array}{ll}
		0 &\mbox{for }0\leq r \leq r_0(m),\\
		\frac{m}{2} \left[r- r_0(m)\right] &\mbox{for }r > r_0(m),
	\end{array}\right.
\end{align}
where $m=\sqrt{{16L_r^2}/{\lambda}}$ is the slope of the solution and $r_0(m)$ is the intersection between the non-zero part of the solution and the $r$-axis, which depends on $m$. In particular, it takes values
\begin{align}
	r_0(m)=2\pi\left(1-\frac{1}{m}\right).
\end{align}
When the condition ${\lambda}/{16 L_r^2}=1$ is satisfied, $m=1$ and $r_0(1)=0$. This way, equation \eqref{SolPezzi} is a general formulation for the condition ${\lambda}/{16 L_r^2}\leq1$ (i.e., $m\geq 1$). It is worth noticing here that this is not the only solution, but other solutions can be defined by alternating the constant parts and non-constant parts in an infinite number of combinations. It would correspond to \textit{pieces} of baryonic layers, leading to a discontinuous form of the energy density. For this reason, we considered only the solution \eqref{SolPezzi}. In this more general case, the energy density takes the form \eqref{rhoEdipA}, but each $\rho_i$ has an additional dependence on the parameter $m$. In particular, for $0\leq r \leq r_0(m)$, the expressions are \footnote{Once again, this is true up to the insertion of brane sources at the boundaries, as required by the Stokes theorem (or, equivalently, by the boundary conditions required to make the variational principle available).}
\begin{widetext}
	{\small 
		\begin{align}
			\rho_0&=K\|c\|^2 \bigg(\frac{p^2}{2L_\phi^2}+\frac{q^2}{4L_\theta^2}\bigg), \label{meaning} \\
			\rho_1&= 0,\\
			\rho_2&=0.
		\end{align}
	}
	On the other hand, for $r>r_0(m)$,
	{\small 
		\begin{align}
			\rho_0&=K\|c\|^2 \bigg\{\frac{p^2}{2L_\phi^2}\left[1+\frac{\lambda m^2}{8L_r^2}+\frac{q^2\lambda}{2L_\theta^2}\sin^2\left(\frac{m(r-r_0(m))}{2}\right)\right]+\frac{q^2}{4L_\theta^2}\left(1+\frac{\lambda m^2}{4L_r^2}\right)+\frac{m^2}{16b^2L_r^2}\bigg\},\\
			\rho_1&= -K\|c\|^2\bigg\{\frac{p^2}{2L_\phi^2}\left[\frac{q\lambda}{L_\theta^2}\sin^2\left(\frac{m(r-r_0(m))}{2}\right)\right]+\frac{q}{L_\theta^2}\sin^2\left(\frac{m(r-r_0(m))}{4}\right)\left(1+\frac{\lambda m^2}{8L_r^2}\right)\bigg\},\\
			\rho_2&=K\|c\|^2\bigg\{\frac{p^2}{2L_\phi^2}\bigg[
			\frac{\lambda}{2L_\theta^2}\sin^2\left(\frac{m(r-r_0(m))}{2}\right)\bigg]+\frac{1}{L_\theta^2}\sin^2\left(\frac{m(r-r_0(m))}{4}\right)\left(1+\frac{\lambda m^2}{8L_r^2}\right)\bigg\},
		\end{align}
	}
\end{widetext}

The baryonic density can be computed in a very similar way, giving 
\begin{align}\label{rhoBSm}
	\rho_B = \left\{\begin{array}{ll}
		0 & \mbox{for } 0\leq r\leq r_0(m)\\
		\frac{\|c\|^2}{8\pi^2}mq\partial_\phi\Phi%
		\sin\left(\frac{m(r-r_0(m))}{2}\right) & \mbox{for } r>r_0(m),
	\end{array}\right.
\end{align}
which integral gives the usual integer value of equation \eqref{LasagnaBaryonicCharge}. Thus, as for the energy density, the baryons accumulate in peaks that become sharper for bigger values of $m$, preserving the baryonic charge $B$. Notice that in the region corresponding to the density \eqref{meaning} there are no baryons, but the energy density is nonzero (and independent of the external field). This energy thus correspond to mesonic fluctuations.

The total energy can be written as
\begin{align}
	E(q;h,m)&=2\pi^2KL_rL_\theta L_\phi\|c\|^2\sum_{i=1}^{4}\Sigma_i(h,m)\ q^i,
\end{align}
where, this time,
{\small
	\begin{align}\label{eta1}
		&\Sigma_4(h,m)=\frac{8L_r^2}{L_\phi^2L_\theta^2}\frac{\pi}{m^3},\\\label{eta2}
		&\Sigma_2(h,m)=\frac{3\pi}{L_\theta^2}+\frac{6\pi}{L_\phi^2}\cr
		&\qquad+h^2\frac{16L_r^2\pi}{L_\phi^2L_\theta^2m^3}\left[\frac{2\pi^2-3}{3m^2}+2\pi^2\left(1-\frac{1}{m}\right)\right],\\\label{eta3}
		&\Sigma_0(h,m)=\frac{\pi m^2}{4L_r^2}\cr
		&\quad +h^2\frac{8\pi}{L_\theta^2m}\left[3\pi^2\left(1-\frac{1}{m}\right)+\frac{1}{m}\left(2-\frac{1}{m}\right)+\frac{\pi^2}{m^2}\right],\\\label{eta4}
		&\Sigma_3(h,m)=-h\frac{16L_r^2\pi^2}{L_\phi^2L_\theta^2m^3}\left(2-\frac{1}{m}\right),\\\label{eta5}
		&\Sigma_1(h,m)=-h\frac{6}{L_\theta^2m}\left[\frac{4}{m}+\pi^2\left(2-\frac{1}{m}\right)\right].
    \end{align}
}
It is straightforward to check that when $m=1$ these quantities reduce to \cref{I,II,III,IV,V}.
The partition function takes the form
\begin{widetext}
\begin{align}
	\mathcal{Z}&=\sum_q\exp\bigg\{-\beta\|c\|^2\bigg[2KL_rL^2\pi^2\sum_{i=1}^{4}\Sigma_i(h,m)q^i-\frac{q^2}{2}\mu_B\bigg]\bigg\},
\end{align}
\end{widetext}
where, again, we used the condition \eqref{Condpq} with $L_\theta=L_\phi=L$. This allows us to determine the thermodynamic properties of the baryonic layers. In this scope, we need to explicitly compute the sums over $q$. This has already been studied for the ungauged case in \cite{Canfora15:CFT}.


\section{The polarization effects}


In this section, we discuss the possibility of considering an electromagnetic field generated from the polarization effects induced by the external field. In this scope, we suppose that to preserve its existence, the skyrmion generates a \textit{backreaction} that cancels the external magnetic field in the boundaries. This gives rise to an internal electromagnetic field, whose behavior can be defined as solutions to the Maxwell Equations.

Let us choose the usual ansatz, with
\begin{gather}
	\alpha=\frac{q}{2}\theta-\frac{p}{2}\left(\frac{t}{L_\phi}-\phi\right),\\
    \xi=\frac{q}{2}\theta+\frac{p}{2}\left(\frac{t}{L_\phi}-\phi\right),\\
    \chi=\chi(r).
\end{gather}
Let us suppose that the induced internal field takes the form
\begin{align}\label{GaugePol}
	A_\mu=\left(-\frac{A_\phi}{L_\phi}(r),0,A_\theta(r),A_\phi(r)\right)^T.
\end{align}
This way, the Skyrme equation for the profile $\chi(r)$ becomes
\begin{widetext}
\begin{align}
	&\partial_{\mu}\partial^{\mu}\chi \left\{1+b^2\lambda\left[%
	\left(\frac{A_\theta^2}{L_\theta^2}-\frac{q}{L_\theta^2}A_\theta\right)\sin\left(\frac{b\chi}{2}%
	\right)+\frac{q^2}{4L_\theta^2}%
	\right]\right\}-b\sin(b\chi)\left(1-\frac{b^2\lambda}{4}\partial_{\mu}\chi%
	\partial^{\mu}\chi\right)\left(\frac{A_\theta^2}{L_\theta^2}-\frac{q}{L_\theta^2}A_\theta\right)=0,
\end{align}
\end{widetext}
Notice that it admits a linear solution
\begin{align}
    \chi(r)=\sqrt{\frac{4L_r^2}{b^2\lambda}}r.
\end{align}
The Maxwell equations for the gauge field \eqref{GaugePol} may be rewritten as follows
\begin{widetext}
\begin{align}
	&\frac{A''_{\theta}}{L_r^2}-\frac{K}{2}\|c\|^2\Bigr\{ \left(\frac{q}{2}-A_{\theta}\right)\Bigr[8\sin^2\left(\frac{\chi}{2}\right)\left(1+\frac{%
		\lambda}{4}\partial_{\nu}\chi\partial^{\nu}\chi\right)-2\lambda\cos^2(\chi)\frac{q^2}{4L_\theta^2}\Bigr]\Bigr\}=0\ ,
\end{align}
\begin{align} 
	&\frac{A''_{\alpha}}{L_r^2}-\frac{K}{2}\|c\|^2\Bigr\{ \left(pq-A_{\alpha}\right)\Bigr[8\sin^2\left(\frac{\chi}{2}\right)\left(1+\frac{%
		\lambda}{4}\partial_{\nu}\chi\partial^{\nu}\chi\right)+2\lambda\sin^2(a\chi)\frac{q^2}{4L_\theta^2}\Bigr]\Bigr\}=0\ ,
\end{align}   
\end{widetext}
where, using the notation of \cite{NostroII},
\begin{align}
    A_{\alpha}=pA_{\theta}+qA_{\phi}.
\end{align}
As discussed in \cite{NostroII}, these equations can be always re-conduced to a Hill equations form. In particular, when the linear solution for $\chi$ is considered, they become Wittaker-Hill equations. Indeed, they may be written as
\begin{gather}
    A_i''+\left[\Lambda_i+\Gamma_i\cos(2\omega)+\Delta_i\cos(4\omega)\right]A_i=0,
\end{gather}
with $i=\theta,\alpha$ and
\begin{gather}
    \Lambda_\theta=\frac{K}{2}\|c\|^2\left(\frac{\lambda q^2}{4L_\theta^2}-4\right),\cr
    \Gamma_\theta=2K\|c\|^2,\quad \Delta_\theta=\frac{K}{2}\|c\|^2\frac{\lambda q^2}{4L_\theta^2},\\
    \Lambda_\alpha=-\frac{K}{2}\|c\|^2\left(\frac{\lambda q^2}{4L_\theta^2}+4\right),\cr
    \Gamma_\theta=2K\|c\|^2,\quad \Delta_\theta=\frac{K}{2}\|c\|^2\frac{\lambda q^2}{4L_\theta^2},\\
    \omega=\frac{\chi}{2}.
\end{gather}

\end{document}